\documentclass[final,3p,12pt,authoryear,times]{elsarticle}
\usepackage{lineno}

\usepackage{graphicx}
\graphicspath{{plots/}}
\usepackage{epsfig,bm,color}
\usepackage{amssymb,amsmath}
\usepackage{natbib}

\usepackage{times}
\usepackage{amsfonts}
\usepackage{multirow}
\usepackage{multicol}
\usepackage{amsmath}
\usepackage{caption}
\usepackage{subcaption}

\journal{Journal}

\begin{document}

\begin{frontmatter}

\title{Assessment of a symmetry-preserving JFNK method for atmospheric convection}

\author[ref2]{M. Alamgir Hossain}
\author[ref1]{Jahrul M Alam\corref{cor1}}
\ead{alamj@mun.ca}
\cortext[cor1]{Corresponding author}

\address[ref1]{Department of Mathematics and Statistics, Memorial University, Elizabeth Ave, NL A1C 5S7, Canada}
\address[ref2]{Department of Mathematics and Statistics, Simon Fraser University, 8888 University Dr, Burnaby, BC V5A 1S6, Canada}

\begin{abstract}
Numerical simulations of nonhydrostatic atmospheric flow, based on linearly decoupled semi-implicit or fully-implicit techniques, usually solve linear systems by a pre-conditioned Krylov method without preserving the skew-symmetry of convective operators. We propose to perform atmospheric simulations in such a fully-implicit manner that the difference operators preserve both the skew-symmetry and the tightly nonlinear coupling of the differential operators. We demonstrate that a symmetry-preserving Jacobian-free Newton-Krylov~(JFNK) method mimics a balance between convective transport and turbulence dissipation. We present a wavelet method as an effective symmetry preserving discretization technique. The symmetry-preserving JFNK method for solving equations of nonhydrostatic atmospheric flows has been examined using two benchmark simulations of penetrative convection -- a) dry thermals rising in a neutrally stratified and stably stratified environment, and b) urban heat island circulations for effects of the surface heat flux $H_0$ varying in the range of $25 \le H_0 \le 930$~W~m$^{-2}$. The results show that an eddy viscosity model provides the necessary dissipation of the subgrid-scale modes, while the symmetry-preserving JFNK method provides the conservation of mass and energy at a satisfactory level. Comparisons of the results from a laboratory experiment of heat island circulation and a field measurement of potential temperature also suggest the modelling accuracy of the present symmetry-preserving JFNK framework.
   
\end{abstract}


\end{frontmatter}

\section{Introduction}
Even with increasing power of computers and advances in numerical methods, it is a challenging endeavour to resolve the important physics of convective motion and cascade of turbulence kinetic energy~(TKE) in atmospheric simulations. The large-scale physics cannot reach a near equilibrium of the interplay between convective transport and diffusive dissipation. We have to cope with the formidable problem of the subgrid-scale parameterization of convective processes~\cite[][]{Ferziger97,Pielke2002,Tannehill2013}.  Mathematically, the convection ( $\bm u\cdot\bm\nabla\bm u$) and the diffusion ($\bm\nabla\cdot\bm\tau$) are governed by skew-symmetric and symmetric positive-definite operators, respectively, which are not fully preserved in many operational atmospheric modelling codes~\cite[see][]{Wicker1998,Skamarock97,Pielke2002}. A parameterization of the subgrid-scale stress $\bm\tau$ would provide a subtle balance of the two operators, which is broken by the non-symmetric discretization of the skew-symmetric convective operator. Most important reasons for preserving the symmetry of convective operator are: i) improved forecasting skill for meso-scale phenomena; and ii) reduced cost for highly complex forecasting systems when unified for both the boundary layer and the meso-scale phenomena. %
As the discretization breaks the skew-symmetry of the convective operator, the underlying conservation law is not globally satisfied at discrete times~\cite[][]{Verstappen2003}. The stability and conservative properties of the existing non-symmetric schemes are thoroughly reviewed by~\cite{Steppeler2003} and~\cite{Klemp2007}. Studies observed that higher-order linearly consistent schemes are sometimes unable to deal with the contamination of poorly resolved short-wavelength perturbations, usually triggered by atmospheric convection~\cite[][]{Skamarock97,Bryan2002,Pielke2002,Steppeler2003}.

In this article, we present the Jacobian-free Newton-Krylov (JFNK) method~\cite[][]{Knoll2004,Zingg2009} for studying mesoscale penetrative convection and thermal dynamics of the atmospheric boundary layer flow~\cite[][]{Carpenter1990,Skamarock97,Bryan2002,Lane2008}. The JFNK method is increasingly considered in many branches of computational fluid dynamics~(CFD). However, it has not been a choice in major atmospheric flow solvers except in a few academic studies~\cite[e.g.][]{Reisner2002,Alam2015}. The lack of a broad acceptance of the JFNK method by the atmospheric science community is somewhat related to the known challenge of constructing appropriate preconditioners. If the nonlinear convection and other physical effects such as turbulence, radiation, or latent heat release are included within the matrix to be inverted at each time step, the construction of a preconditioned JFNK method for atmospheric modelling is not fully clear from the existing literature. The present study fills in the research gap in atmospheric modelling, where we demonstrate that preserving the symmetry of operators in their discretization can partially circumvent the preconditioning challenge through a physics-based nonlinear preconditioning approach. 

To develop a JFNK solver, we discretize the tightly coupled residual of mass, momentum, and energy of the nonhydrostatic atmospheric model equations~\cite[see][]{Bryan2002} in a nonlinearly consistent manner. To account for the lack of implicit dissipation by the skew-symmetric discretization of the convective operator, we show that a subgrid model is capable of dissipating the short-wavelength perturbations triggered by atmospheric convection. For the proposed JFNK method, we consider a wavelet-based approximation of differential operators, which filters the short-wavelength perturbation. We study penetrative convection and convective boundary layer (CBL) flow over a heterogeneously heated surface. Although convection may not illustrate all of the computational issues of atmospheric modelling, the study of thermal dynamics indicates that the JFNK method offers much insight into the more complicated dynamics of atmospheric convection. We demonstrate that a rising thermal penetrating into a stably stratified atmosphere will eventually overshoot its level of neutral buoyancy, a crucial component of which is the generation of internal gravity waves. This overshooting involves entertainment and detrainment, which plays a key role in atmospheric mixing and convective redistribution of heat and other scalars. Capturing such phenomena of atmospheric convection illustrates our understanding of the JFNK methodology in dealing with the coupled nature of atmospheric multiphysics and the fascinating nonlinear cascade of scales of atmospheric dynamics.

Section~\ref{sec:wmdl} presents the JFNK methodology for solving the governing equations for compressible nonhydrostatic atmospheric flows, where a technical details of the wavelet-based discretization is outlined briefly. Section~\ref{sec:nra} summarizes the numerical results of penetrative turbulent convection in the atmospheric boundary layer. We have discussed the results with respect to neutrally- and stably-stratified configurations. The test cases considered in this article are representative cases for the verification of atmospheric modelling. Results of other numerical models and field measurements have been utilized to validate the symmetry preserving JFNK methodology. Finally, Section~\ref{sec:sfd} discusses the present findings and outlines how the presented methodology may further be extended to advance the field of atmospheric modelling.

\section{Methodology}\label{sec:wmdl}
\subsection{Governing equations}
Let us consider the dynamics of idealized dry thermals without condensation, evaporation, or any background wind shear, where the continuity, momentum, and energy equations take the following form in Cartesian coordinates~\cite[see][]{Skamarock97,Pielke2002,Bryan2002},
 \begin{equation}
 \frac{\partial p}{\partial t} + u_j \frac{\partial p}{\partial x_j} = - \frac{c_p}{c_v} \left(\frac{\partial u_i}{\partial x_i}\right)p, 
\label{eq:ch2_28}
\end{equation}
\begin{equation}
\frac{\partial u_i}{\partial t} +  u_j \frac{\partial u_i}{\partial x_j} = - \theta_0 \frac{\partial p}{\partial x_i} - \theta_0 \left(\frac{\partial p_0}{\partial x} \delta_{i1}+\frac{\partial p_0}{\partial y} \delta_{i2} \right) + \frac{\partial \tau_{ij}}{\partial x_j} +  \frac{\theta}{\theta_0} g \delta_{i3},%
\label{eq:ch2_29}
\end{equation}
\begin{equation} 
\frac{\partial \theta}{\partial t} + u_j \frac{\partial \theta}{\partial x_j}+ u_j\delta_{j3} \beta = \frac{\partial\tau_{\theta j}}{\partial x_j}.    
\label{eq:ch2_30}
\end{equation}
The equations~(\ref{eq:ch2_28}-\ref{eq:ch2_30}) are nonlinearly coupled by the convective operator. The conservative properties and stability of this model are directly related to the energy contribution and the propagation speed of atmospheric waves~\cite[][]{Steppeler2003}. The velocity $u_i$ is coupled with the non-dimensional pressure ($p$) that is referred to as the Exner function and related to the dimensional pressure $P$:
\[p = c_p\left( \frac{P}{p_0}\right)^{R/c_p}. \]
In Eq~(\ref{eq:ch2_30}), a splitting the total potential temperature is considered, such as $\theta_0 + \bar\theta + \theta$, where $\theta_0$ is a constant background temperature,  $\beta = \partial \bar \theta/\partial z$, $\bar\theta(z)$ denotes a mean vertical distribution of the temperature, and $\theta$ is a temperature perturbation. This decomposition is often useful for implementing the heat flux boundary condition at the ground~\cite[][]{Dubois2009}. Note, $p_0 = 1000$ hPa is the reference pressure, $R = 287$ J kg$^{-1}$ K$^{-1}$ is the gas constant,  $c_p = 1004$ J kg$^{-1}$ K$^{-1}$ is the specific heat at constant pressure, and $c_v = 717$~J~kg$^{-1}$ K$^{-1}$ specific heat at constant volume. In Eqs~(\ref{eq:ch2_28}-\ref{eq:ch2_30}),  $x_i$  denotes the Cartesian coordinate $\bm x = (x,y,z)$, $\delta_{ij}$ is the Kronecker delta, $\tau_{ij}$ is the turbulent momentum flux, and $\tau_{\theta j}$ is the turbulent heat flux. Note that $u_i$ and $u_j$ denotes the velocity components $u_1,\,u_2,\,u_2$; however, as mentioned below, the bold-face $\bm u = [u_i, p, \theta]$ represents the numerical solution vector of the system~(\ref{eq:ch2_28}-\ref{eq:ch2_30}). 

\subsection{Symmetry preserving discretization}\label{sec:trb}
To illustrate how the symmetry of underlying physics is preserved numerically by the JFNK method, let $\bm u = [u_k]$ be a column vector $[u_i,p,\theta]_k$, {\em i.e.} the numerical solution of Eqs~(\ref{eq:ch2_28}-\ref{eq:ch2_30}) at each spatial grid point $\bm x_k,\,\hbox{for }\, k=0\ldots\mathcal N-1$, and $\mathcal F$ represent the discretization of all spatial differential operators involved in the system~(\ref{eq:ch2_28}-\ref{eq:ch2_30}). Then, the following dynamical system
\begin{equation}
  \label{eq:ds}
  \frac{\partial\bm u}{\partial t} = \mathcal F\bm u
\end{equation}
represents the spatially discretized form of Eqs~(\ref{eq:ch2_28}-\ref{eq:ch2_30}).
Considering a time centred implicit (trapezoidal) scheme for the dynamical system~(\ref{eq:ds}), we get
\[\frac{2\bm u^{n+1}}{\Delta t}-\mathcal F (\bm u^{n+1})=\frac{2\bm u^{n}}{\Delta t}+\mathcal F (\bm u^{n}), \]
which is a nonlinear system of algebraic equations of the following compact form
\begin{equation}
\mathcal{L}(\bm u^{n+1})=\bm f(\bm u^n).
\label{eq:ns}
\end{equation}
For a nonlinear problem, the time centred scheme~(\ref{eq:ns}) with a fixed positive step size $\Delta t$ leads to a bounded error for $n\rightarrow\infty$, which is equivalent to A-stability of the scheme when it is applied to a linear problem. Since the order of an A-stable linear multistep method cannot exceed $2$, the time centered method is the best choice to deal with waves not contributing to energy conservation in the solutions of  atmospheric model equations~(\ref{eq:ch2_28}-\ref{eq:ch2_30})~\cite[see, e.g.][]{Randall90}.

The nonlinear convective operator $u_j\partial u_i/\partial (\cdot)$ is skew-symmetric because of the property of the trilinear form that
$\langle u_j\partial v_i,w_i\rangle + \langle w_i, u_j\partial v_i\rangle$.
In Eq~(\ref{eq:ds}), the operator $\mathcal F$ is said to be skew-symmetric with respect to the inner product $\langle\cdot\rangle$  if we have $\langle\mathcal F\bm u, \bm u\rangle + \langle\bm u, \mathcal F\bm u\rangle=0$ for all vectors $\bm u$. In other words, an anti self-adjoint operator is skew-symmetric. If  the operator $\mathcal F$ is a matrix, the skew-symmetry is equivalent to $\mathcal F=-\mathcal F^T$.
Now, taking the inner product of $\bm u$ with both the sides of Eq~(\ref{eq:ds}) and ignoring the effects of boundary conditions, we find that
$$
\frac{\partial}{\partial t}\langle\bm u,\bm u\rangle = \langle\mathcal F\bm u,\bm u\rangle + \langle\bm u,\mathcal F\bm u\rangle.
$$
Clearly, the dynamical system~(\ref{eq:ds}) conserves the inner product $\langle\bm u,\bm u\rangle$ if the corresponding operator $\mathcal F$ satisfies the above skew-symmetric property. In order to satisfy the conservation of the inner product $\langle\bm u,\bm u\rangle$  at discrete level in the context of the dynamical system~(\ref{eq:ds}),  we must have the inner product satisfying $\langle\bm u^{n+1} ,\bm u^{n+1}\rangle = \langle\bm u^{n} ,\bm u^{n}\rangle$ for two consecutive time steps. It can be shown that such a requirement at each time step is satisfied, subject to a truncation error of $\mathcal O(\Delta t^2)$, by the trapezoidal time integration scheme~(\ref{eq:ns}) considered above if the operator $\mathcal F$ is skew-symmetric. 

Consider a higher-order upwind-biased discretization of convection in Eq~(\ref{eq:ds}), which minimizes the local truncation error. An upwind discretization does not retain the skew-symmetry of the convective operator. With classical upwind methods, the eigenvalues of the operator $\mathcal F$ will have negative real parts~\cite[see][]{Klemp2007}. The negative real part of the eigenvalues of the operator $\mathcal F$ help ensure the stability of the system~Eq~(\ref{eq:ds}). However, they artificially dampen the energy $\langle\bm u,\bm u\rangle$, and thus, the conservation of energy cannot be satisfied globally~\cite[see][]{Klemp2007}.
To preserve the skew-symmetry of the convective differential operator $u_j\partial u_i/\partial x_j$ with a second order finite difference method, momentum transport equation can be discretized in the following skew-symmetric form
$$
\mathcal F\bm u = \frac12\frac{\partial u_i u_j}{\partial x_j} + \frac12 u_j\frac{\partial u_i}{\partial x_j},
$$
where one considers the arithmetic mean of flux- and convective-forms of the convective operator.

Notice that the skew-symmetry of the convective operator is directly related to the conservation of the convective variable. It is worth mentioning that the nonlinear stability of numerical schemes is often easier to establish if the nonlinear convective term is expressed in the skew-symmetric form. 
First, preserving skew-symmetry results in reduced levels of artificial dissipation, which is desired in atmospheric simulations. Second, it eliminates the convective instability associated with spurious transfer of kinetic energy on grids that are not fine enough to resolve short-wavelength perturbations caused by convective transport. For example, it was reported by previous researchers that due to  the artificial numerical damping of the shorter-wavelength, the Weather Research and Forecasting (WRF-ARW) model is unable to adequately resolve the capping inversions.  Third, it ensures that the numerical dissipation of resolved kinetic energy does not overwhelm the dissipation provided by a subgrid-scale model.

Preserving the skew-symmetry of the convective transport by the wavelet method, in addition to the tightly nonlinear coupling of the JFNK method, brings multifold benefits discussed above.

\subsection{Wavelet-based collocation method}
\begin{figure}
  \centering
  \begin{tabular}{cc}
    $(a)$ & $(b)$\\
    \includegraphics[height=6cm]{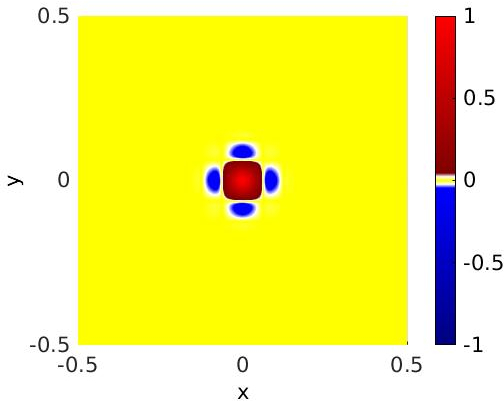}
    &
      \includegraphics[height=6cm]{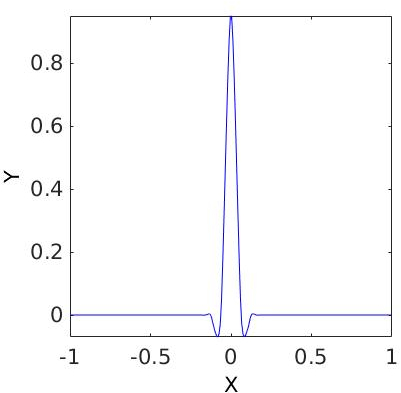}
  \end{tabular}
  \caption({$a$) A wavelet function $\psi_k(\bm x)$ satisfying $\psi_k(\bm x_l)=\delta_{kl}$; it takes a value of $1$ on a given grid point $\bm x_k\in\mathcal G^j$ and $0$ on all other grid points $\bm x_l\in\mathcal G^j$. Then, $\psi_k(\bm x)$ is extended to all grid points $\bm x_k\in\mathcal G^{j+1}$ by~\cite{Deslauriers1989} interpolation, and subsequent iterations forms a continuous function in $\mathcal G^j$ as $j\rightarrow\infty$. $(b)$ A restriction of $\psi_k(\bm x)$ on $y=0$ is displayed to indicate the symmetry and support of $\psi_k(\bm x)$.}
  \label{fig:psi}
\end{figure}

Deslaurier-Dubuc interpolating wavelets~\cite[see][]{Deslauriers1989,Mallat2009} are defined on a sequence of nested grids
$
\mathcal G^j = \{\bm x_k \}
$
which are embedded over nested approximation spaces $\mathcal V^j\subset\mathcal V^{j+1}$. An element of the basis $\{\psi_k(\bm x)\}$  of $\mathcal V^j$ is presented in Fig~\ref{fig:psi}$a,b$. The wavelet collocation method finds an approximation $u^{\mathcal N}(\bm x)\in\mathcal V^j$ of $u(\bm x)\in L^2(\Omega)$ such that
$
\langle\mathcal L(u^{\mathcal N}) - f,\tilde\psi_k\rangle = 0,
$
where $\mathcal N$ is the number of grid points, $\mathcal L$ is a differential operator including the boundary conditions, and $\{\tilde\psi_k\}$ is a dual basis corresponding to the approximation space $\mathcal V^j$. For the given basis $\{\psi_k(\bm x)\}$ of $\mathcal V^j$, there exist a dual approximation space equipped with a basis $\{\tilde\psi_k(\bm x)\}$ such that $\tilde\psi_j(\psi_i)=\delta_{ij}$. 

The wavelet-based approximation %
\begin{equation}
  \label{eq:mwd}
  \begin{array}{lll}
    u^{\mathcal N}(\bm x) %
    &=& \sum\limits_{k=0}^{\mathcal N-1}u(\bm x_k)\psi_k(\bm x)\\
  \end{array}
\end{equation}
projects the coefficients $\{u(\bm x_k)\}$ into $\mathcal V^j$ in which the projection $u^{\mathcal N}(\bm x)$ does not oscillate at wavelengths smaller than the grid-spacing. 
The discretization of differential operators are performed through projection of derivatives into $\mathcal V^j$. Without the details of wavelet theory, the wavelet-based projection~\cite[see][for a technical details]{Deslauriers1989} of the first derivative with respect to $x$ is given by
$$
\frac{\partial}{\partial x} u^{\mathcal N}(\bm x) = \sum\limits_{k=0}^{\mathcal N} u'(\bm x_k)\psi_k(\bm x) = \sum\limits_{k=j-2p+1}^{j+2p-1}u(\bm x_k)\frac{\partial}{\partial x}\psi_k(\bm x).
$$
The symmetry of differential operators are preserved due to the symmetry of $\psi_k(\bm x)$. On a uniformly refined grid having a grid-spacing of $\Delta x$ in all directions, the local truncation error is $\mathcal O(\Delta x^{2p})$ for the above approximation of derivatives~\cite[][]{Alam2014}. The derivative of $u(\bm x)$ is exactly represented by this wavelet method if $u(\bm x)$ is a polynomial of degree $2p-1$. The subgrid-scale modes of half the wavelength of the resolved scale modes, which are contributed by convection $u\partial u/\partial x$ contributes, are explicitly filtered and parameterized by the subgrid model. 

\subsection{The subgrid scale closure model}\label{sec:sgs}
In atmospheric modelling~\cite[see][]{Pielke2002}, the sugrid-scale schemes assume that turbulence produces vertical mixing in the real atmosphere, and that the role of the horizontal components of the subgrid scale stress is to control nonlinear aliasing errors~\cite[][]{Pielke2002}.
Such schemes are based on the momentum exchange coefficient~\cite[e.g.][]{dear70,Deardorff80},%
$$
\nu_{\tau} = \left(\Delta_{\hbox{\tiny LES}}C_s\right)^2\sqrt{2\mathcal S_{ij}\mathcal S_{ij}},
$$
where the stresses%
$$
\tau_{ij} = 2\nu_{\tau}\mathcal S_{ij} - \frac{1}{3}\tau_{kk}\delta_{ij}
$$
are related to the strain $\mathcal S_{ij} = \frac{1}{2}\left(\frac{\partial u_i}{\partial x_j} + \frac{\partial u_j}{\partial x_i}\right)$ of the resolved flow.  Computing resources limit atmospheric simulations on coarse grids, where the subgrid model acts on turbulent motions that are anisotropic and intermittent~\cite[][]{Moeng2015}. Moreover, the vertical dissipation remains stronger than the horizontal dissipation, for example in penetrative convection~\cite[][]{Bartello2013}. To examine the symmetry preserving JFNK solver with respect to a basic subgrid model, we follow the dimensional reasoning outlined by~\cite{dear70} (see Eq~3.4 therein) to estimate the horizontal and the vertical exchange coefficients separately,
\begin{equation}
  \label{eq:KmM}
  K_m = C_s^{4/3}~\varepsilon^{1/3}~ (\Delta_x\Delta_y)^{2/3} \quad\hbox{and }\quad  K_M = C_s^{4/3}~\varepsilon^{1/3}~ \Delta_z^{4/3}.
\end{equation}
Here,  $C_s$ is a dimensionless constant and the rate of dissipation of turbulent kinetic energy is $\varepsilon = \nu_{\tau}\mathcal S_{ij}\mathcal S_{ij}$. According to~\cite{Deardorff80}, the horizontal eddy diffusivity is $K_h = (1+2l/\sqrt{\Delta_x\Delta_y})K_m$ and the vertical eddy diffusivity is $K_H = (1+2l/\Delta_z)K_M$, where $l=0.75\sqrt{\tau_{kk}}/N$ is a subgrid scale mixing length and $N$ is the Brunt-V\"{a}is\"{a}l\"{a} frequency.

\subsection{A brief outline of the JFNK method}\label{sec:jfnk}
To ensure a minimal technical details of present contribution, we closely follow the preconditioned Krylov method considered by~\cite{Skamarock97} in their semi-implicit scheme for solving the linearized nonhydrostatic atmospheric model Eqs~(\ref{eq:ch2_28}-\ref{eq:ch2_30}). 
The JFNK method is a class of practical iterative methods for finding the solution $\bm u^*$ to the nonlinear system $\mathcal L(\bm u) = \bm f$, Eq~(\ref{eq:ns}), when an initial approximation, $\bm u^{0}$, is known. The nonlinear function $\mathcal L:\mathbb R^{\mathcal N}\rightarrow\mathbb R^{\mathcal N}$ is assumed differentiable and $\mathcal J(\bm u)$ denotes the Jacobian of $\mathcal L$ at the point $\bm u$. %

\subsubsection{Convergence rate of Newton-Krylov solvers}
For the nonlinear system~(\ref{eq:ns}), let us find variations $\delta\bm u^k$ of the solution vector $\bm u^{n+1}$ iteratively such that $\bm u^{n+1,k+1} = \bm u^{n+1,k} + \delta\bm u^k$ for $k\ge 0$. This outer loop of  iterations forms the Newton's method. The solution from the previous time step provides the first iteration $\bm u^{n+1,0}=\bm u^{n}$. At $k$-th Newton iteration, we minimize the residual vector $\mathcal R(\bm u^{n+1,k})\, \dot{=}\,  \mathcal L(\bm u^{n+1,k}) - f(\bm u^n)$ using the generalized minimal residual~(GMRES) method of~\cite{Saad1986}. Thus, we look for the variation $\delta\bm u^k$ satisfying the linear system
\begin{equation}
  \label{eq:jv}
  \mathcal J(\bm u)\delta\bm u^k = -\mathcal R(\bm u^{n+1,k}). %
\end{equation}
Only a few Krlov iterations of the inner loop solve Eq~(\ref{eq:jv}). This `inexact' Newton-Krylov method is equivalent to solving the ordinary differential equation
$$
\frac{d\bm u^{n+1,k}}{dk} = -\mathcal J^{-1}(\bm u)\mathcal R(\bm u^{n+1,k})
$$
by the Euler-explicit method with a step size of one. Therefore, $\mathcal R(\bm u^{n+1,k}) = e^{-k}\mathcal R(\bm u^n)$ if Eq~(\ref{eq:jv}) is solved exactly. This shows the fast rate of convergence of the inexact Newton's method.

\subsubsection{Physics-based nonlinear preconditioning}
Two families of preconditioning method are usually considered for the JFNK method. The linear preconditioning is quite similar to the Krylov method presented by~\cite{Skamarock97}. For solving the linear system~(\ref{eq:jv}) by a preconditioned Krylov method, linear preconditioning is classified as right- and left-preconditioning. In contrast, a physics-based nonlinear preconditioning is cost-effective thanks the wavelet-based discetization. Each diagonal block of the operator $\mathcal L$ in Eq~(\ref{eq:ns}) means a physical field that couples with itself and each off-diagonal block means a physical field that couples with another field. Consider a physics-based preconditioner in which the weak coupling of off-diagonal terms in the operator $\mathcal L$ is ignored. Physics-based preconditioner gathers the eigenvalues of the preconditioned system in small areas, thereby increasing the convergence rate.

The physics-based nonlinear preconditioning approach constructs an equivalent system of non-linear equations which provides faster rate of convergence with respect to the original system. It can be shown that if a fixed point iteration converges for the system~(\ref{eq:ns}), the eigenvalues of the Jacobian $\tilde{\mathcal J}(\bm u) = \partial\tilde{\mathcal R}(\bm u)/\partial\bm u$ for the preconditioned nonlinear system $\tilde{\mathcal R}(\bm u) = \bm u - \mathcal P^{-1}(\bm u) [\bm f(\bm u^n) - \mathcal L(\bm u) + \mathcal P(\bm u)]$ are gathered in a small area, where $\mathcal P(\bm u)$ is the nonlinear preconditioning matrix. The most attractive feature of nonlinear preconditioning is that faster convergence rate of Krylov iteration is achieved with minimal mathematical and coding effort. For example, in case of implementing the JFNK method within an existing atmospheric modelling code, a fixed point iteration can be performed with only a few code modification. 

For the matrix-vector product on the left side of Eq~(\ref{eq:jv}), the JFNK method needs to compute the action of the linear map $\mathcal J:\mathcal V^l\subset\mathbb R^{\mathcal N}\rightarrow\mathbb R^{\mathcal N}$ on the variation $\delta\bm u^k$ of the solution vector $\bm u^{n+1}$. To compute this action with a complexity of $\mathcal O(\mathcal N)$, consider the Fr\'{e}chet derivative of the operator $\mathcal R(\bm u)$ defined by
\begin{equation}
  \label{eq:fd}
  \mathcal J(\bm u)\delta\bm u^k = \lim_{\eta\rightarrow 0}\frac{\partial}{\partial\eta}\mathcal R(\bm u + \eta\delta\bm u^k).
\end{equation}
The limit in Eq~(\ref{eq:fd}) exists, and thus, $\mathcal R(\bm u)$ is Fr\'{e}chet differentiable, where
\[
  \lim_{\eta\rightarrow 0}\frac{||\mathcal L(\bm u^{n+1,k}+\eta\delta\bm u^k) - \mathcal L(\bm u^{n+1,k}) - \mathcal J\delta\bm u^k||}{||\eta\delta\bm u^k||} = 0.
\]
In other words, the same algorithm that provides the differentiation matrix $\mathcal L$ is applied to calculate the action of $\mathcal J$ on $\delta\bm u^k$ without requiring additional technical development a preconditioner. This observation suggest that the implementation of JFNK method within an existing atmospheric modelling code is straight forward. Moreover, the complexity of the JFNK method scales like the complexity of the algorithm used for the discretization of Eqs~(\ref{eq:ch2_28}-\ref{eq:ch2_30}), which is $\mathcal O(\mathcal N)$ for the wavelet method.
\section{Numerical results and discussion}\label{sec:nra}
We report primary results on the accuracy, efficiency, and efficacy of the symmetry-preserving JFNK method as a potential candidate for problems of meteorological interest. We have studied two categories of convective phenomena to test the tightly nonlinear strategy of the JFNK method. 
Comparisons of the present results among experimental and numerical data collected from the literature have been considered. These  numerical exercises indicate that the tightly nonlinear physics-based coupling of all physical processes considered within the JFNK method has the potential to be a scale-adaptive frame-work for modelling the transition from the near-surface small-scale 3D physics to the outer-layer meso-scale meteorology. 
\subsection{Penetrative convection and rising thermals: reference model \label{refmodel}}
We have compared the JFNK simulation of penetrative convection with the results provided by~\cite{Bryan2002}, \cite{Wicker1998}, and \cite{Carpenter1990}. %
\cite{Bryan2002} examined a time-split segregated algorithm in which the convective operator was discretized in its flux-form without preserving its skew-symmetry. They adopted a divergence damping term to help maintain the quality of the scheme.
 \cite{Carpenter1990} noted that the choice of not preserving skew-symmetry of convective operators by the positive definite upwind schemes is to help control the Gibbs phenomenon. They also reported upwind schemes tend to smear sharp gradient of penetrative thermals. %

We consider that a warm perturbation of $\theta$ is prescribed at the horizontal midpoint and at a height of $2$~km in the domain of $[-10,10]\times[0,10]\hbox{ km}^2$, where the surrounding environment is neutrally stratified with lapse rate of $10^{\hbox{\tiny o}}$~K~km$^{-1}$ and $\theta_0=300$~K. The initial thermal has a radius of $2$~km~\cite[e.g.][]{Bryan2002}. As mentioned in Table~\ref{tab:khrare}, momentum- and heat-exchange coefficients are varied for Case A at a fixed Prandtl number of $\mathcal Pr=0.71$. In Case~B, the Prandtl number is varied between $0.5$ and $2.0$. This test confirms hypothesis that short-wavelength modes, triggered by the nonlinear convection process in a period of time evolution of the thermal, are accurately filtered by the subgrid model.

\begin{table*}[t]
\begin{center}
\begin{tabular}{|cccccc|}
\hline\hline
 & $K_M$ (m$^2$ s$^{-1}$) & $K_H$ (m$^2$ s$^{-1}$) &~ $\mathrm{Re}$ ~& ~ $\mathrm{Ra}$~ &~~~ $\mathrm{Pr}$~~~\\
\hline
\multirow{3}{*}{Case A} &10 &  14.1   &  $2.5 \times 10^3$  & $4.44 \times 10^6$& 0.71  \\ 
  & 5 & 7.04  &  $5.0 \times 10^3$   & $1.78 \times 10^7$& 0.71 \\
  & 2.5 &  3.52 & $1.0 \times 10^4$ & $ 7.1 \times 10^7$&0.71\\ 
  & 1.0 & 1.41 & $2.5 \times 10^4$ & $4.44 \times 10^8$&0.71\\
 \hline
\multirow{3}{*}{Case B} & 10 &  5  & $2.5 \times 10^3$ &  $4.44 \times 10^6$ & 0.5 \\ 
 & 5 & 5    & $5.0 \times 10^3$ &  $1.78 \times 10^7$ & 1.0 \\
 & 2.5 &  5 & $1.0 \times 10^4$ &  $7.1 \times 10^7$& 2.0 \\ 
\hline

\end{tabular}
\end{center}
\caption {List of representative parameters for the simulation of two cases in a neutral environment. In case A, the  Rayleigh number $\mathcal Ra$ is varied with fixed Prandtl number $\mathcal Pr=K_M/K_H$. In case B, both the  Rayleigh number $\mathcal Ra$ and the Prandtl number $\mathcal Pr$ are varied. In both cases, the bulk Richardson number is fixed at $\mathcal Ri_b = 0.1 $. The corresponding Reynolds number $\mathcal Re$ is listed in this table.}\label{tab:khrare}
\end{table*}

\begin{table*}[t]
\begin{center}
\begin{tabular}{|lrrrrr|}
\hline
$K_M$ & $10$ m$^2$ s$^{-1}$&  $5$ m$^2$ s$^{-1}$ & $2.5$ m$^2$ s$^{-1}$  & $1.0$ m$^2$ s$^{-1}$ & B \& F\\
\hline
$\theta_{\min}$ (K) &  -0.000632   &  -0.003814  & -0.009359 & -0.133971& -0.144409\\ 
$\theta_{\max}$ (K) & 1.408749  &  1.629635   &   1.843659 &  2.138108 & 2.02178 \\
$u_{\min}$ (m s$^{-1}$) &  -9.511412 & -10.058257 &  -10.636190 & -11.667357 & \\ 
$u_{\max}$ (m s$^{-1}$) & 9.512040 &  10.059020 &   10.637147 &  11.668235& \\ 
$w_{\min}$ (m s$^{-1}$) &  -6.360285  & -6.527770 & -6.596165 & -6.627753 & -8.58069\\ 
$w_{\max}$ (m s$^{-1}$) &  15.352078  & 15.599483 &   15.833058 & 16.018170 &  14.5396\\
$\omega_{\min}$ (s$^{-1}$) & -0.065906 & -0.095523 & -0.137061 & -0.188188 & \\
$\omega_{\max}$ (s$^{-1}$) &  0.065906 &  0.095523 & 0.137049 & 0.188187 & \\ 
\hline
\end{tabular}
\end{center}
\caption {Max/min values of $\theta$, $u$, $w$, and $\omega$ for the dry thermal simulation (Case A) at $t = 1\,000$~[s]. The reference data from~\cite{Bryan2002} is in the column denoted by `B \& F'.} 
\label{tab:extreme_values_ch4_1}
\end{table*}

\begin{table}[t]
\begin{center}
\begin{tabular}{|lrrr|}
\hline 
$K_M$ & $10$ m$^2$ s$^{-1}$& $ 5 $ m$^2$ s$^{-1}$ & $2.5$ m$^2$ s$^{-1}$\\
\hline 
$\theta_{\min}$ (K) & -0.0061   &   -0.0062  &  -0.0063 \\ 
$\theta_{\max}$ (K) &  1.7138   &  1.7268   &  1.7371 \\
$u_{\min}$ (m s$^{-1}$) &  -9.7735 & -10.1486 &  -10.4817\\ 
$u_{\max}$ (m s$^{-1}$) & 9.7744 &   10.1494 &   10.4824\\ 
$w_{\min}$ (m s$^{-1}$) &  -6.5080  & -6.5204 &  -6.5481\\ 
$w_{\max}$ (m s$^{-1}$) &  15.4442  & 15.6334 &   15.8046 \\
$\omega_{\min}$ (s$^{-1}$) & -0.0849 & -0.1053 &  -0.1271 \\
$\omega_{\max}$ (s$^{-1}$) &  0.0849 & 0.1053  & 0.1271 \\ 
\hline
\end{tabular}
\end{center}
\caption {Max/min values of $\theta$, $u$, $w$, and $\omega$ for the dry thermal simulation (Case B, see Table~\ref{tab:khrare}), where $\theta_0 = 300$ K and $t = 1\,000$ s.} 
\label{tab:extreme_values_ch4_2}
\end{table}
 
\begin{table}[t]
\begin{center}
\begin{tabular}{|ccccc|}
\hline
  ~~ $\mathcal Ri_b$~~ &~~ $\mathcal Fr$~~ &~~~~ $N$ (s$^{-1}$)~~~~ & $\omega/N$ & $\alpha$\\
\hline
 1.0  & 1.0 &  $2.5 \times 10^{-2}$ & $0.999$ &$2.65^{\hbox{\tiny o}}$\\
 0.25  & 2.0 & $1.25 \times 10^{-2}$ & $0.997$ & $4.44^{\hbox{\tiny o}}$\\
 0.2  & 2.24 & $1.12 \times 10^{-2}$ & $0.985$ & $9.94^{\hbox{\tiny o}}$\\ 
 0.16 & 2.5  & $1.0 \times 10^{-2}$ & $0.977$ & $12.3^{\hbox{\tiny o}}$\\
0.1  & 3.16 & $7.9 \times 10^{-3}$ & $0.962$ & $15.8^{\hbox{\tiny o}}$\\
0.05 &4.47 & $5.6 \times 10^{-3}$ & $0.929$ & $21.7^{\hbox{\tiny o}}$\\ 
\hline
\end{tabular}
\end{center}
\caption {Relation between the wave frequency and buoyancy frequency for penetrative convection in a stably stratified environment.} 
\label{tab:khrare1}
\end{table}

\subsubsection{Thermals in a neutral environment}\label{sec:ntrl}
\begin{figure}
  \centering
  \begin{tabular}{cc}
    $(a)$ & $(b)$\\
    & \\
    \includegraphics[width=19pc,angle=0]{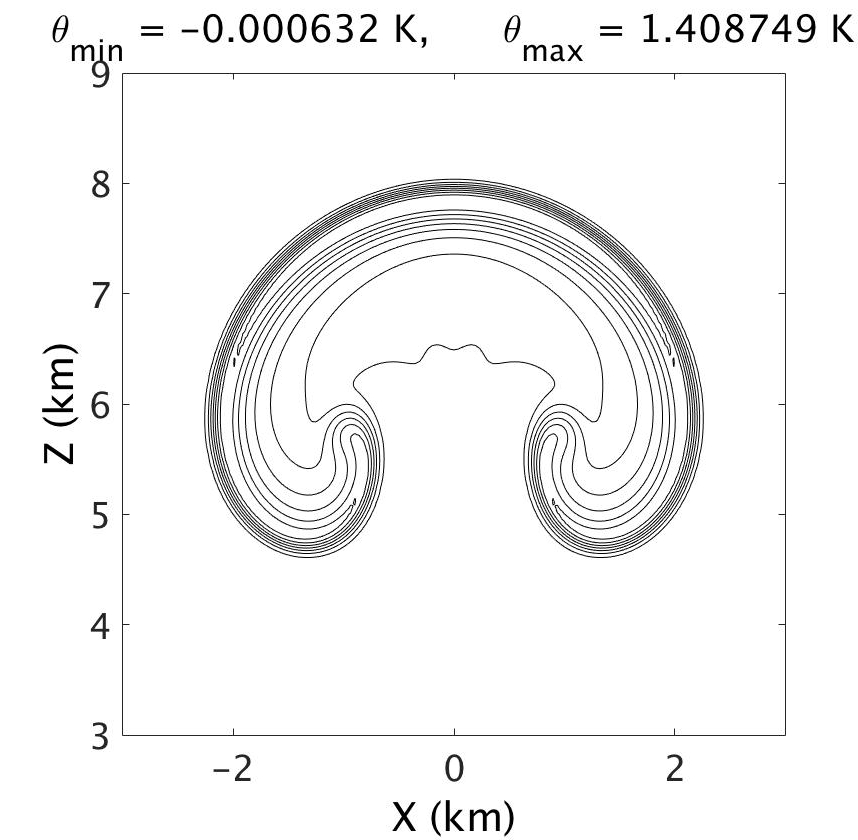}
    &
      \includegraphics[width=19pc,angle=0]{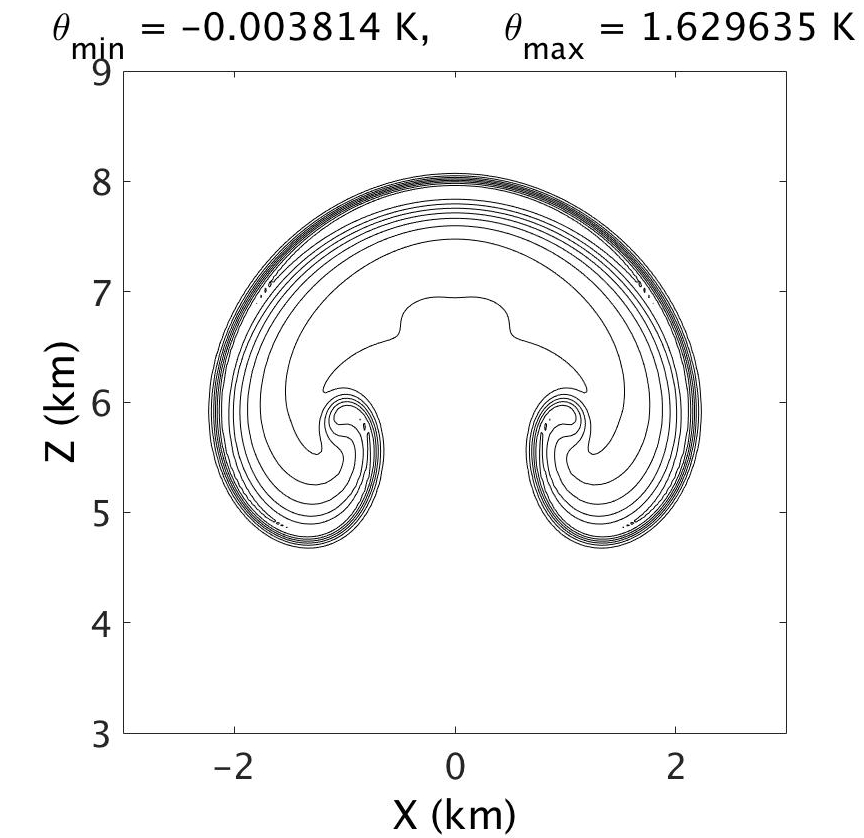}\\
    & \\
    \multicolumn{2}{c}{$(c)$}\\
    & \\
    \multicolumn{2}{c}{\includegraphics[width=19pc,angle=0]{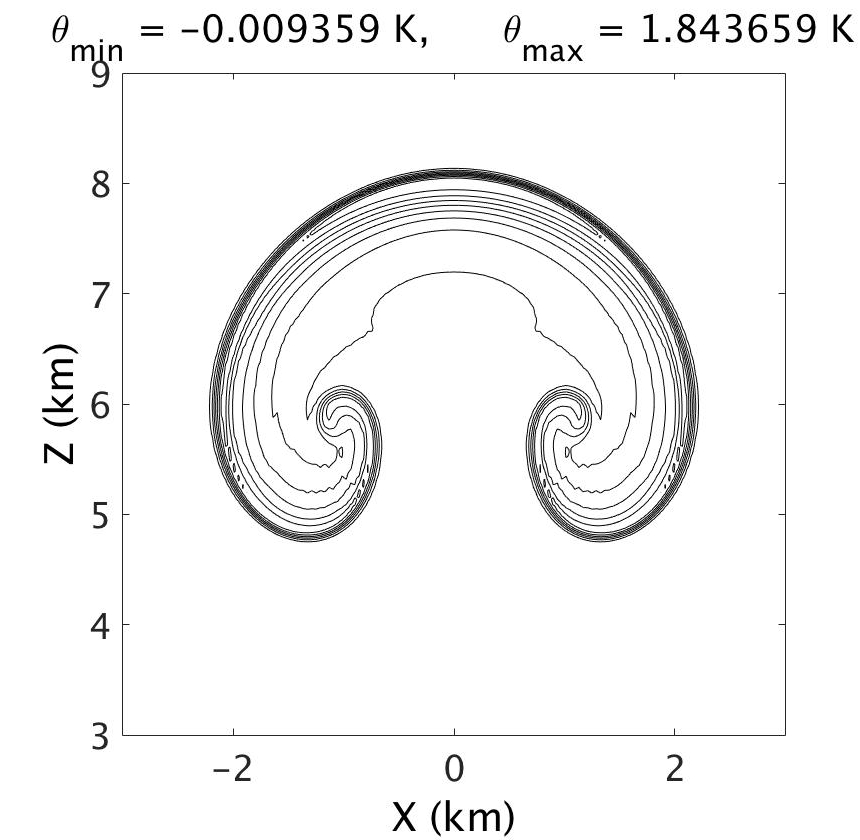}}\\
  \end{tabular}
 \caption{Contour plots of the potential temperature perturbation $(\theta)$  for Case A with a neutral environment at $\mathcal Pr= 0.71$ and  $t=1\,000$ s;  (a) $K_M = 10$ m$^2$ s$^{-1}$; (b) $K_M = 5$ m$^2$ s$^{-1}$, and (c) $K_M = 2.5$ m$^2$ s$^{-1}$.\label{fig:thetacontour_dry}}
\end{figure}

In Case A-B, the effect of the horizontal momentum exchange coefficient, {\em e.g.} $K_M = 10,~5,~2.5,$ and~$1.0$~m$^2$~s$^{-1}$, with respect to the skew-symmetry of convective operator is studied for atmospheric convection in a neutral environment ({\em i.e.} the Buoyancy frequency $N = 0$). The contour plots of the potential temperature $\theta$ in Fig~\ref{fig:thetacontour_dry} shows  the development of two `rotors' around the rising thermal, which replicates the corresponding dynamics predicted by the mesoscale models of~\cite{Wicker1998} and~\cite{Bryan2002}. In comparison to Fig~1 of~\cite{Bryan2002}, one finds that the dynamics of penetrative thermals in a neutrally stratified dry atmosphere has been accurately simulated by the tightly nonlinear coupling strategy of the JFNK method. In particular, the sensitivity of the simulated dynamics on the values of $K_M$ (Table~\ref{tab:khrare}) is consistent with the similar results that appeared in the literature~\cite[e.g.][]{Carpenter1990}. For a quantitative comparison, we note that the minimum and maximum potential temperature reported by~\cite{Bryan2002} are $\theta_{min} = -0.144409$ K and $\theta_{max} = 2.07178$ K, respectively. Table~\ref{tab:extreme_values_ch4_1} indicates a good agreement of the present results with the corresponding values reported by~\cite{Bryan2002}, subject to the differences in the subgrid model and the truncation error of the numerical scheme.

It is worth mentioning that the atmospheric modelling community adopts the upwind scheme for accurate numerical predictions of weather events. Clearly, the symmetry-preserving JFNK method provides numerical predictions of equivalent accuracy. Table~\ref{tab:extreme_values_ch4_1} indicates that the potential temperature field is predicted relatively accurately by the JFNK method for the smallest of the considered values of $K_M=1.0$~m$^2$s$^{-1}$. However, the vertical velocity is predicted more accurately with a higher value of  $K_M=10.0$~m$^2$s$^{-1}$. It is also evident from Table~\ref{tab:extreme_values_ch4_1} that the numerical predictions are not noticeably sensitive to changes in momentum exchange coefficients. 
One observes that the upper surfaces of thermals at $t=1\,000$~s (in Fig~\ref{fig:thetacontour_dry}) are located at heights of $8.04,~8.08,~\hbox{and }8.15$~km for $K_M = 10, ~5,~\hbox{and }2.5$ m$^2$ s$^{-1}$, respectively. %
From the predicted maximum vertical velocity in Table~\ref{tab:extreme_values_ch4_1}, we see that the rate of vertical momentum transfer in a turbulent penetration of dry thermals may be weakly sensitive to the subgrid-scale mixing length $l$ provided by the eddy diffusivity model of~\cite{Deardorff80}. 

\begin{figure}
  \centering
  \begin{tabular}{cc}
    $(a)$ & $(b)$\\
    \includegraphics[trim=0.5cm 1cm 1cm 1cm,clip=true, width=19pc,angle=0]{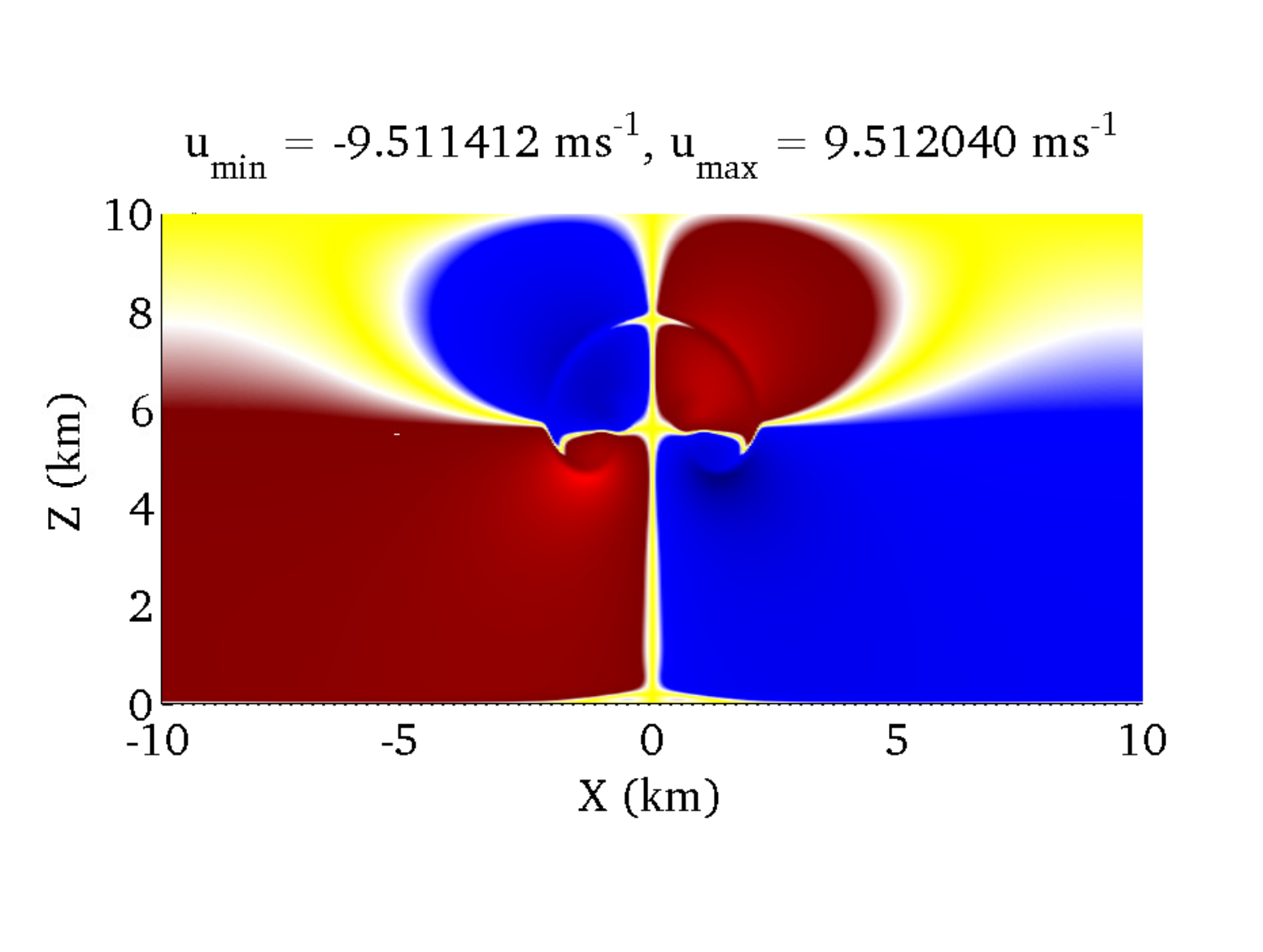}
    &
      \includegraphics[trim=0.5cm 1cm 1cm 1cm,clip=true, width=19pc,angle=0]{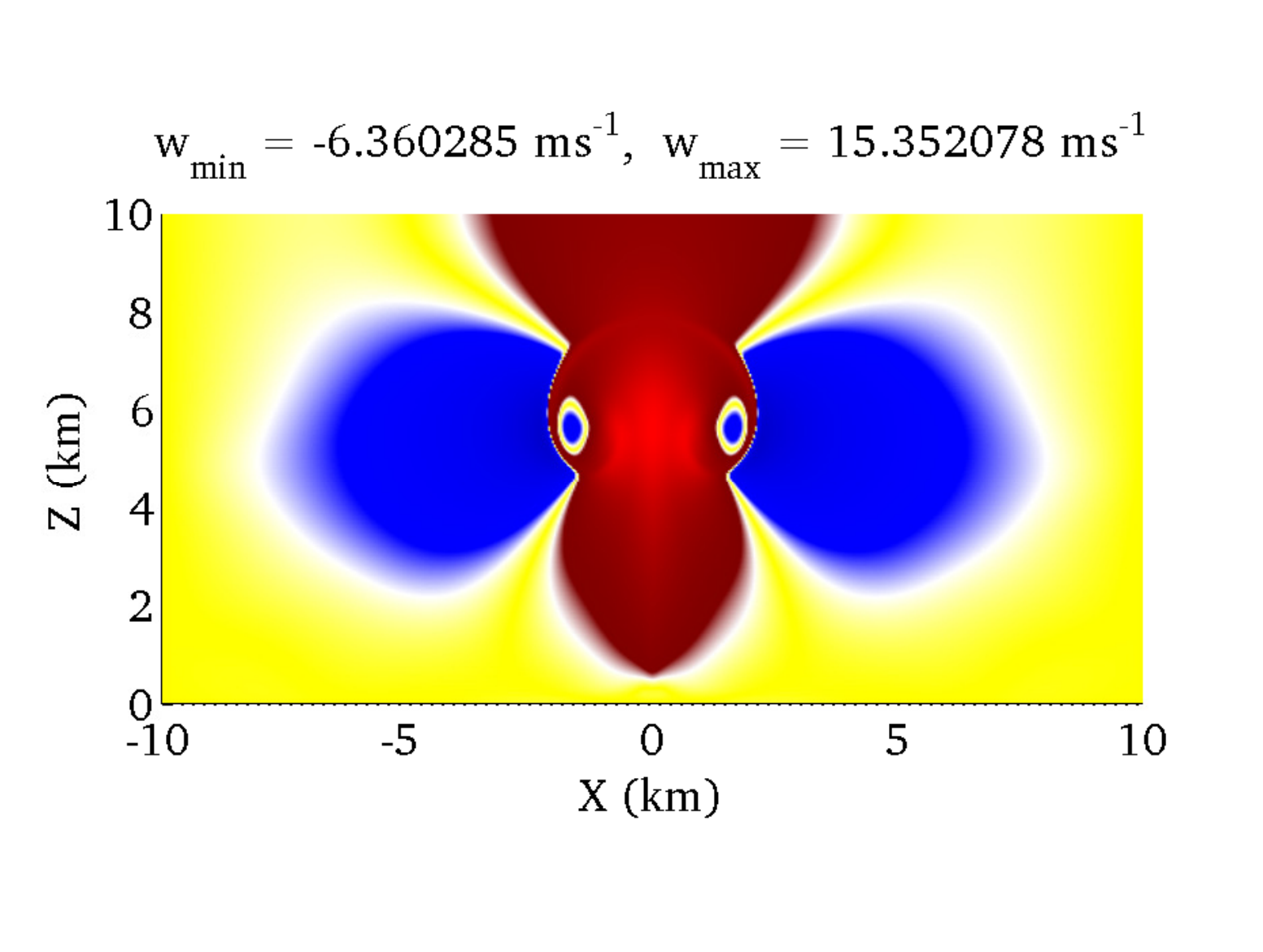}\\
    $(c)$ & $(d)$\\
    \includegraphics[trim=0.5cm 1cm 1cm 1cm,clip=true, width=19pc,angle=0]{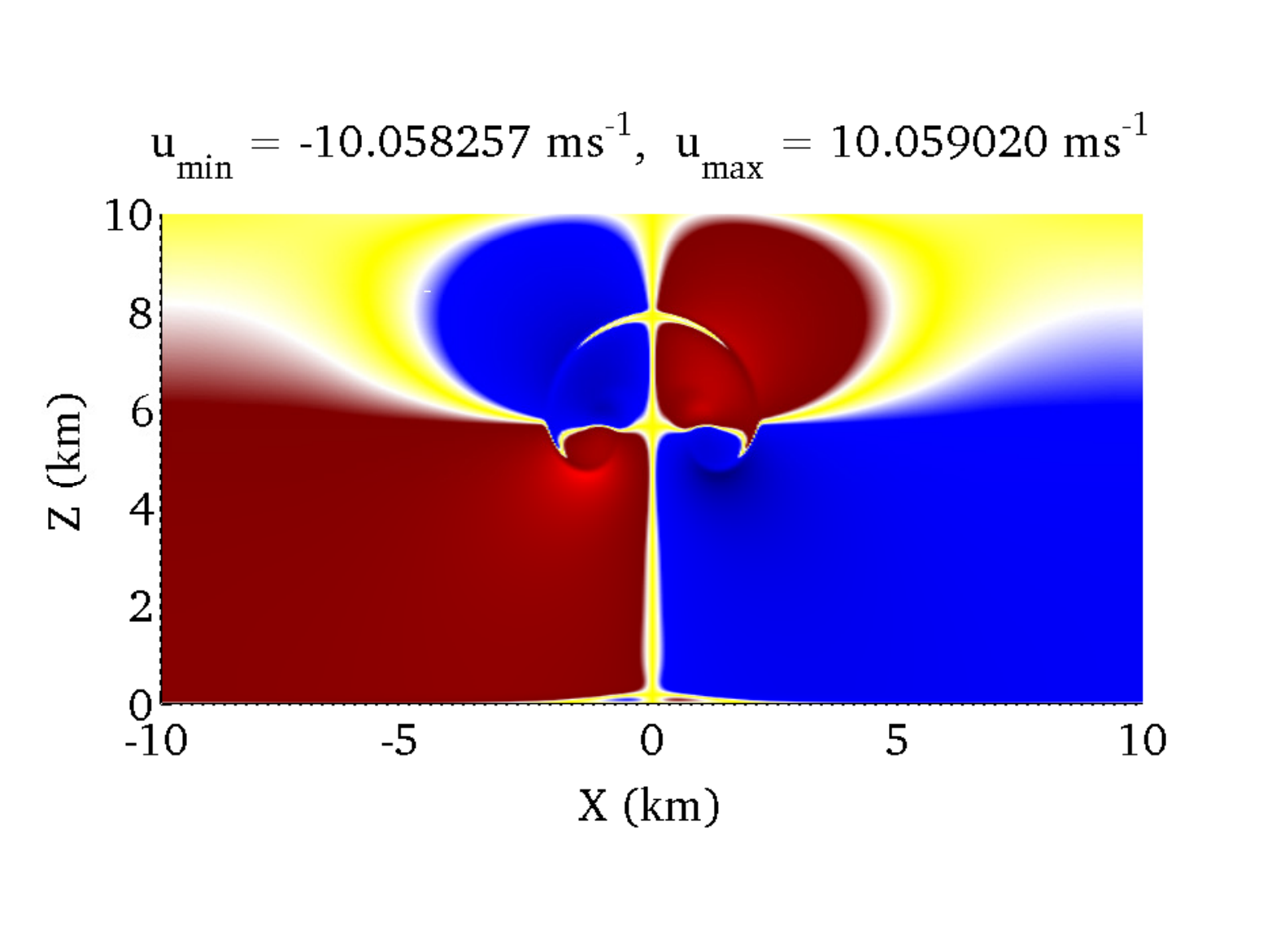}
    &
      \includegraphics[trim=0.5cm 1cm 1cm 1cm,clip=true, width=19pc,angle=0]{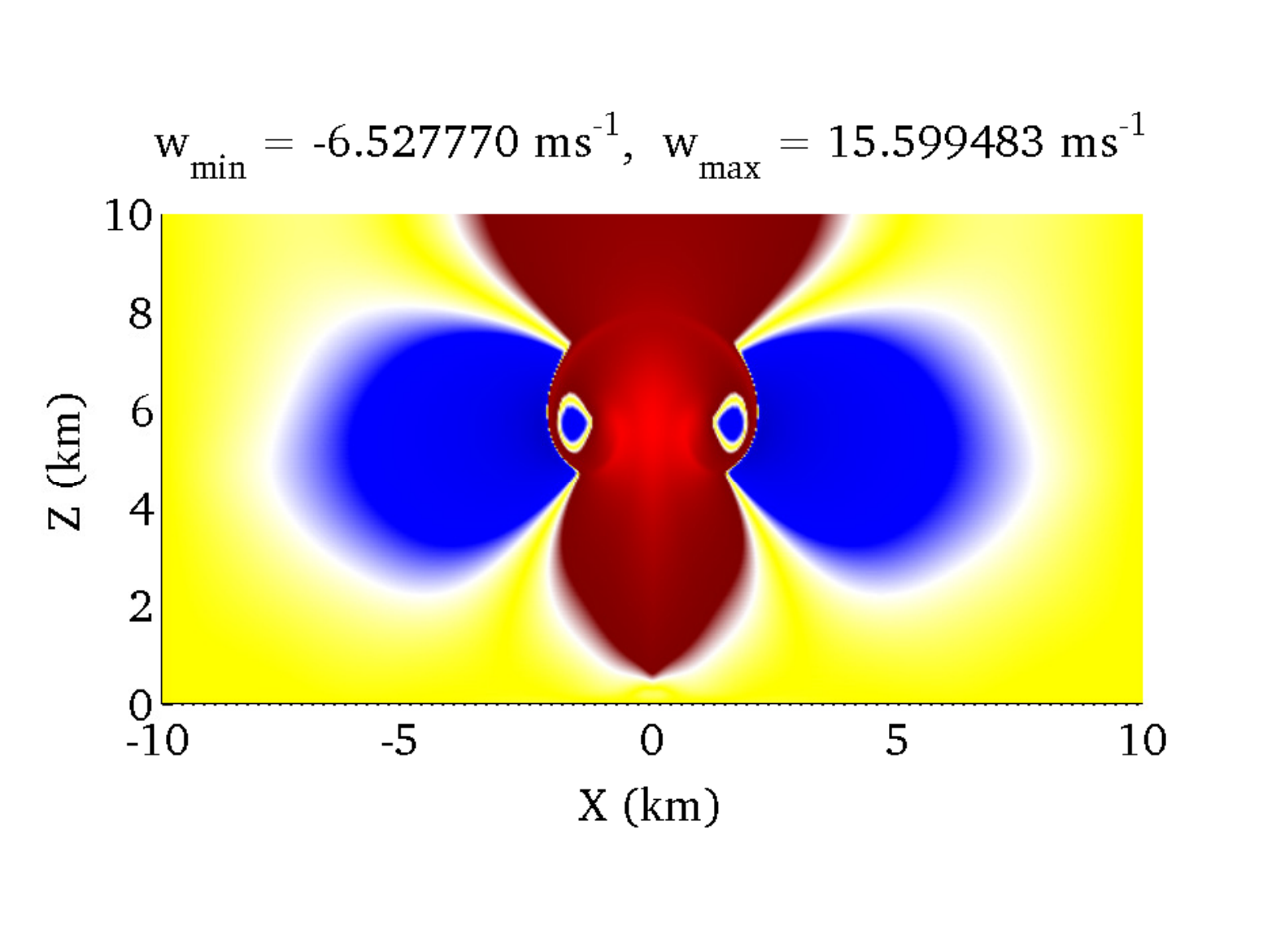}\\
    $(e)$ & $(f)$\\
    \includegraphics[trim=0.5cm 1cm 1cm 1cm,clip=true, width=19pc,angle=0]{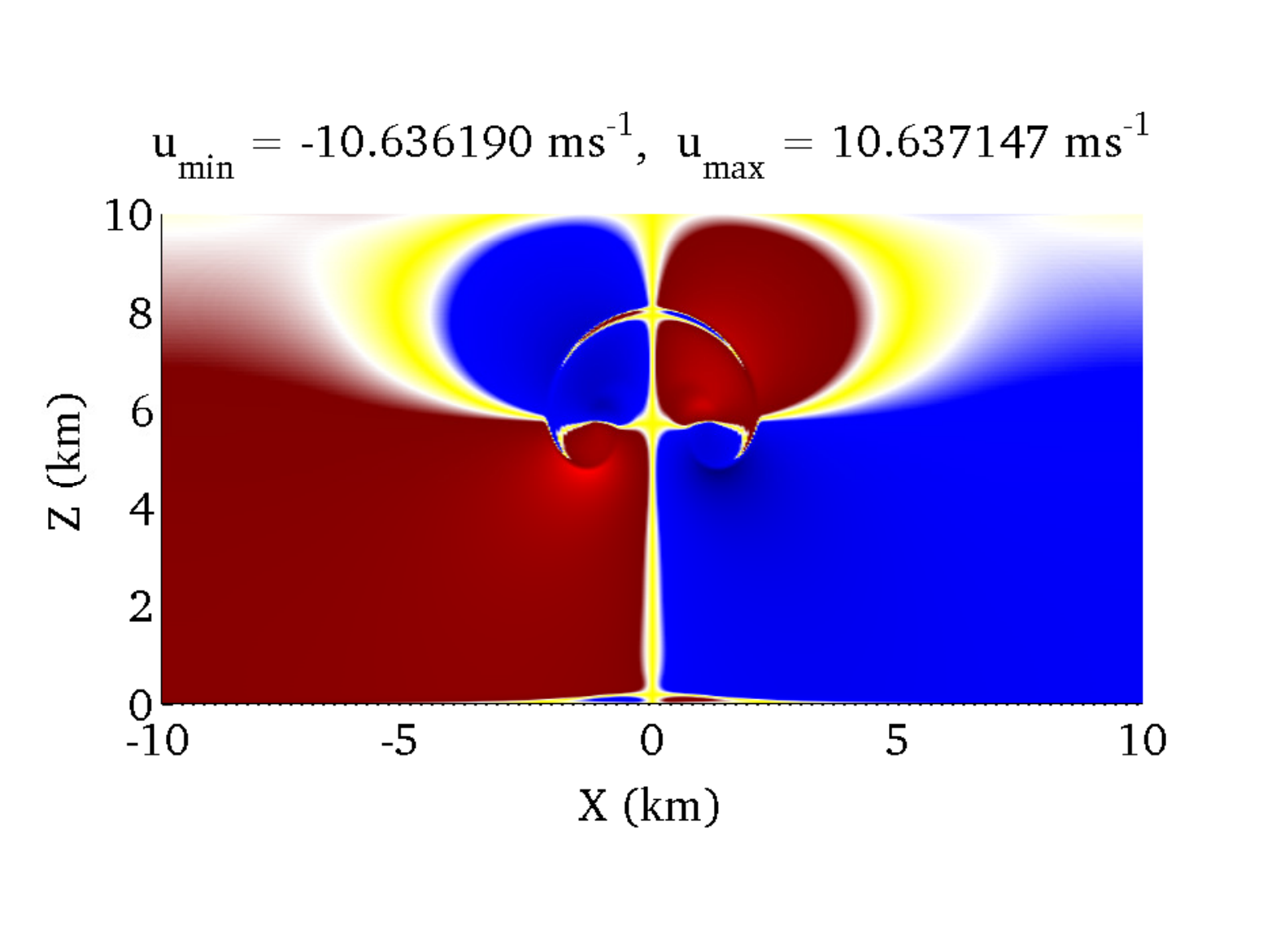}
    &
      \includegraphics[trim=0.5cm 1cm 1cm 1cm,clip=true, width=19pc,angle=0]{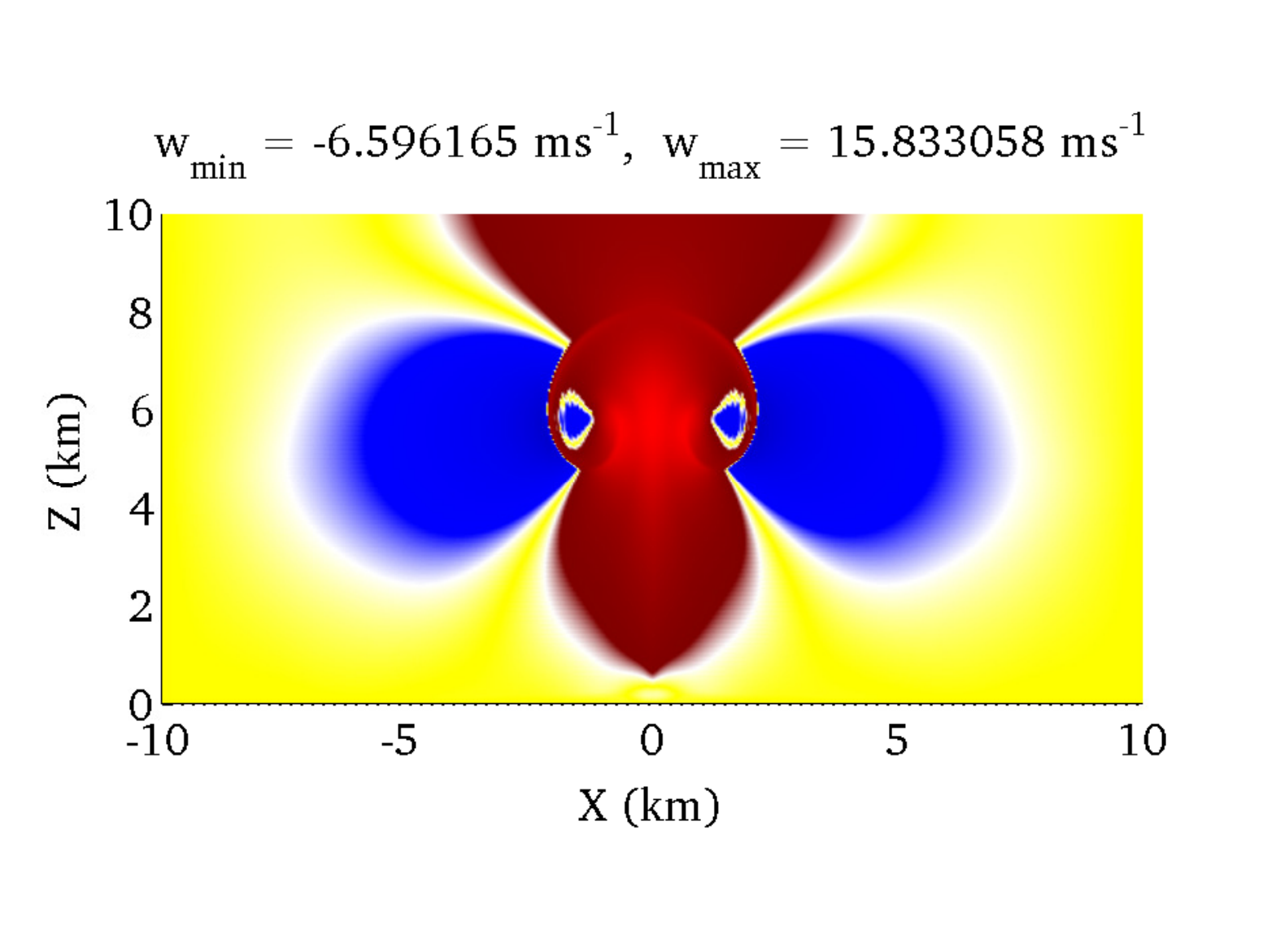}\\
  \end{tabular}
 \caption{Color filled contour plots of the horizontal velocity ($u$, left column) and the vertical velocity ($w$, right column) for Case A with a neutral environment at  $t=1\,000$ s.  (a,b) $K_M = 10$ m$^2$ s$^{-1}$, (c,d) $K_M = 5$ m$^2$ s$^{-1}$ and (e,f) $K_M = 2.5$ m$^2$ s$^{-1}$.  {Red, blue and yellow colors represent positive, negative and zero values, respectively.} \label{fig:velocitycontour_neutral}}
\end{figure}

Color-filled contour plots of the horizontal and the vertical velocities are presented in Figure~\ref{fig:velocitycontour_neutral}. Notice that the velocity field is symmetric about $x=0$~\cite[see also][]{Lane2008}. 
These plots show that a reduction of the subgrid scale momentum-exchange coefficient by $50\%$ does not introduce a significant overall variation of the vertical velocity. The results for Case B (see Table~\ref{tab:khrare}), where the value of $\mathcal Pr$ was also varied, are similar to that for Case A, as reported in Table~\ref{tab:extreme_values_ch4_2}.

\subsubsection{Thermals in a stably stratified environment}
Corresponding to the neutral simulations (Case-A and Case-B), penetrative convection in a stably stratified environment is simulated, where the model was initialized for the potential temperature $\theta_0 + \beta z + \theta(x,z)$. In this situation, the frequency of internal wave ($\omega$) correlates with the buoyancy frequency ($N$) through the dispersion relation $\omega = N\cos\alpha$. Values of the buoyancy frequency $N$ and $\alpha$ are listed in Table~\ref{tab:khrare1}.

\begin{figure}[htbp]
  \centering
  \begin{tabular}{c}
    \includegraphics[width=11cm]{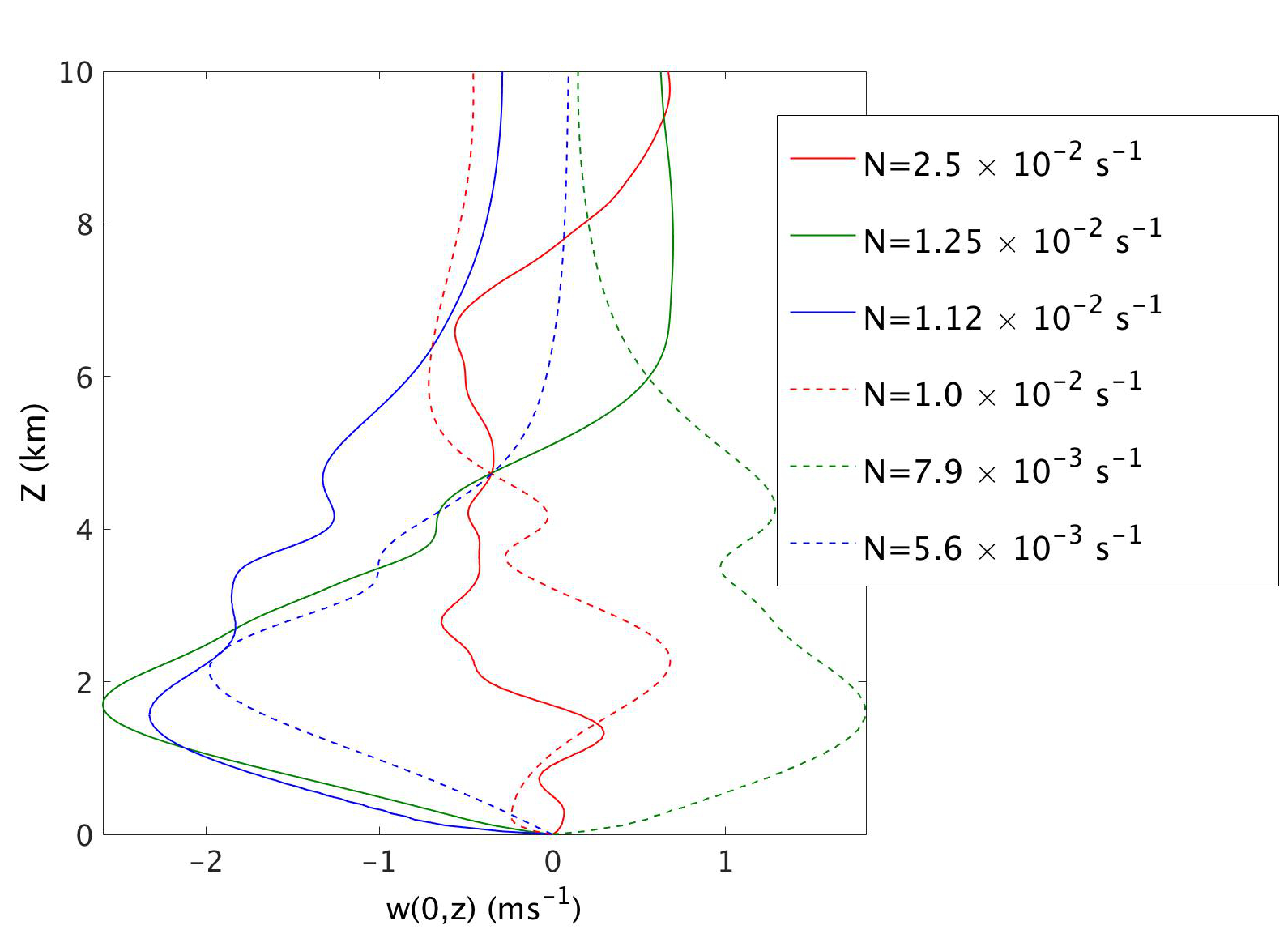}
    \\
    $(a)$ vertical velocity\\
    \includegraphics[width=11cm]{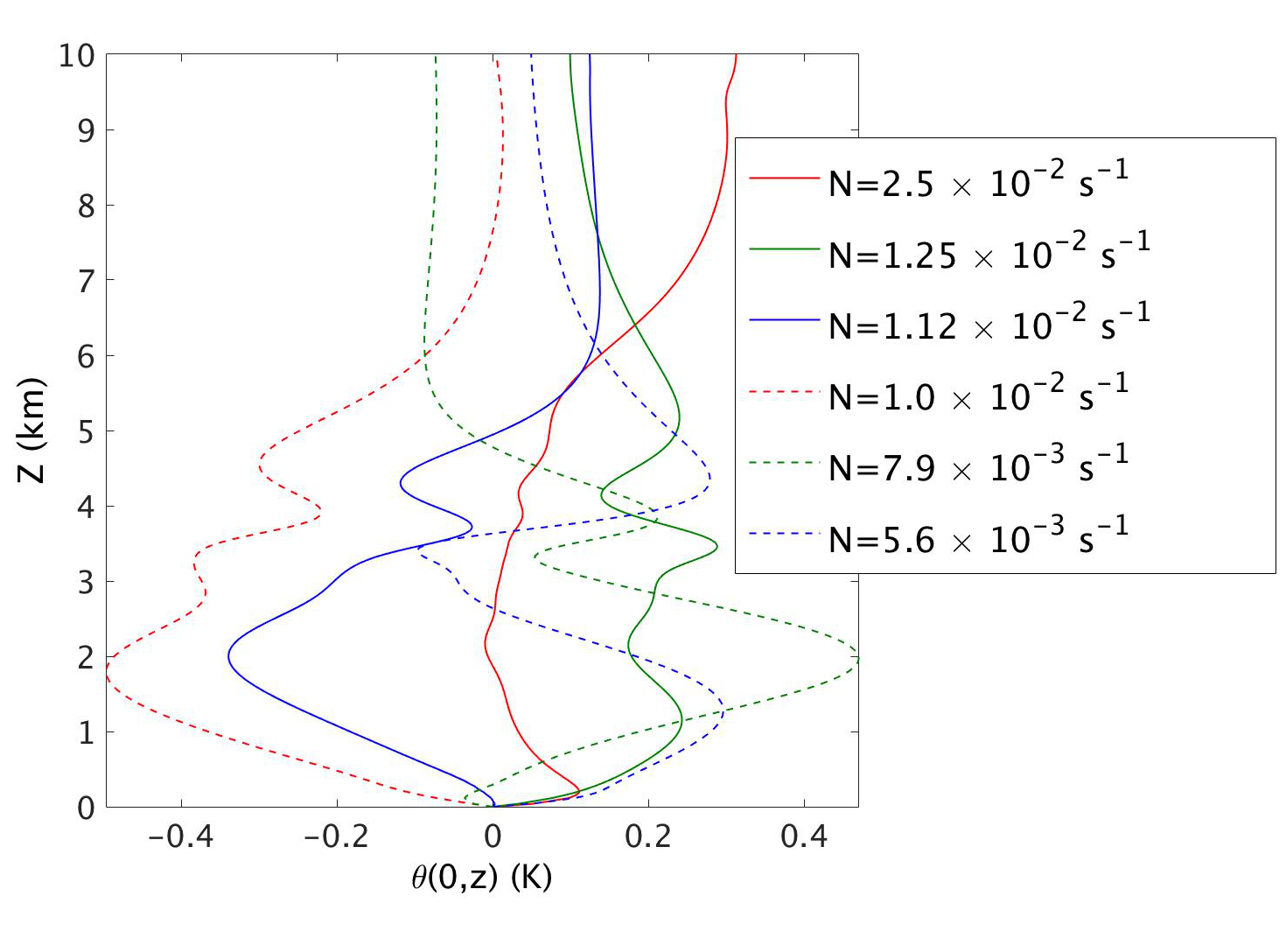}
    \\
    $(b)$ perturbation potential temperature
  \end{tabular}
 \caption{Plots of (a) vertical velocity (b) potential temperature probed along the vertical line $x = 0$ at $t = 1\,000$ s, where the thermal penetrates into a stably stratified environment. Note that $K_M = 10$ m$^2$ s$^{-1}$ and $\mathcal Pr=0.71$.\label{fig:yslice_dristra}}
\end{figure}

The linear theory suggests the existence of evanescent and vertically propagating waves~\cite[see][p. 187]{Lin2007} if %
\[ \frac{2\pi}{N} > \frac{L}{U}\quad \hbox{and}\quad \frac{2\pi}{N} < \frac{L}{U},\] respectively.
Moreover, if $\frac{2\pi}{N} >> \frac{L}{U}$, the buoyancy force becomes extremely weak so that the vertical velocity can be estimated by~\cite[see][]{Lin2007}
\[ w(x,z) = W(x) e^{-k |z|}.\]
Here,  $U$ and $W$ are horizontal and vertical velocity scales, respectively, and $L$ denotes a horizontal length scale. 
Simulated vertical velocity $w(0,z,t)$ and potential temperature $\theta(0,z,t)$ at $t = 1\,000$~s are shown in Fig~\ref{fig:yslice_dristra}. 
The absolute maximum of the vertical velocity and of the potential temperature appear around at $2$~km from the ground. It is also evident that the rising thermals in the stable environment features overshooting that reaches up to a height of $5$ km.

\begin{figure}
 \centering             
        \begin{subfigure}[b]{0.42\textwidth}
          \centering                      
          \includegraphics[width=17.5pc,height=17pc,angle=0]{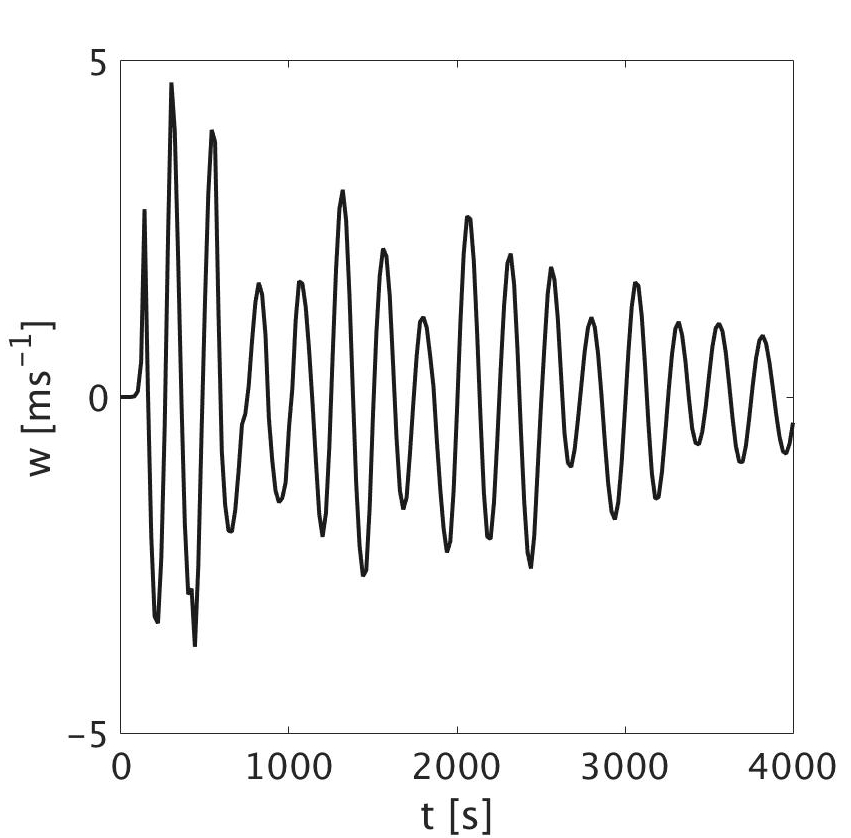}
                \caption{$N = 2.5 \times 10^{-2}$ s$^{-1}$}
                \label{fig:time_series_w_G1_Ri1}
        \end{subfigure}
        ~ %
        \begin{subfigure}[b]{0.43\textwidth}
          \centering                      
                \includegraphics[width=18pc,angle=0]{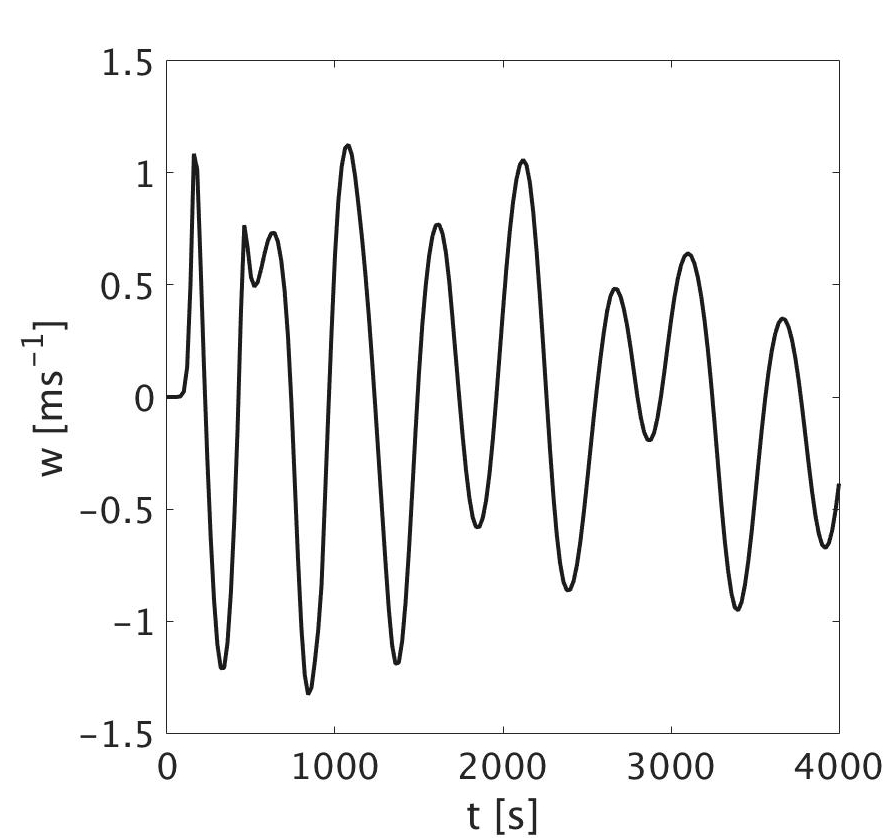}
                \caption{$N = 1.25 \times 10^{-2}$ s$^{-1}$}
                \label{fig:time_series_w_G1_Ri_25}
        \end{subfigure}

        \begin{subfigure}[b]{0.45\textwidth}
          \centering                      
                \includegraphics[width=18pc,angle=0]{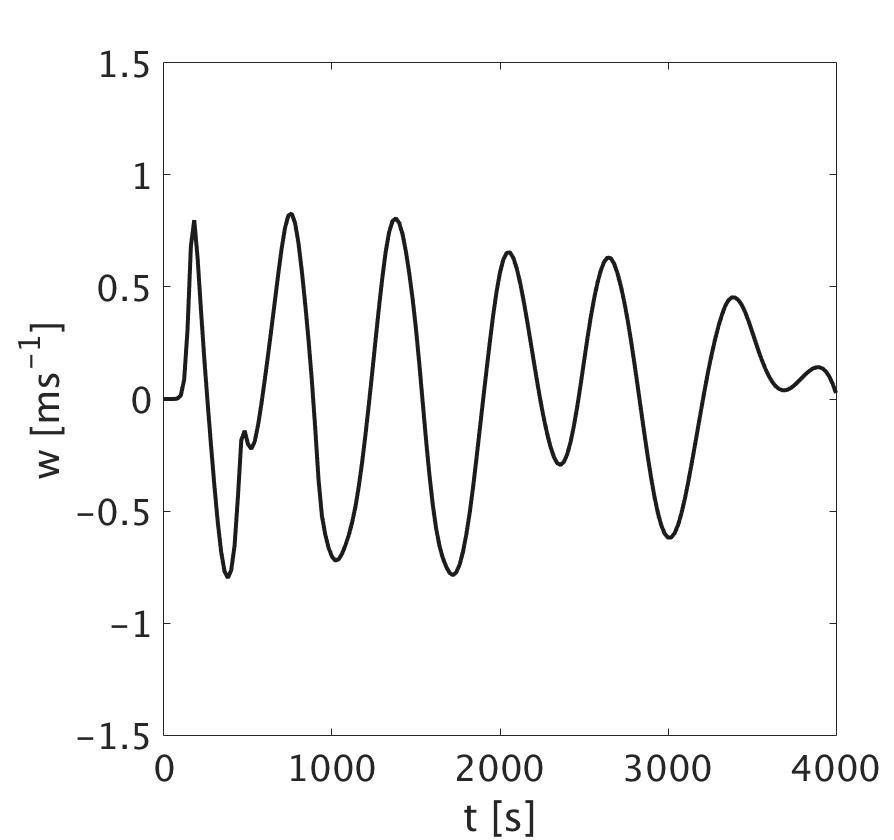}
                \caption{$N = 1.0 \times 10^{-2}$ s$^{-1}$}
                \label{fig:time_series_w_G1_Ri_16}
          \end{subfigure}
           \begin{subfigure}[b]{0.45\textwidth}
                \centering                      
                \includegraphics[width=18pc,angle=0]{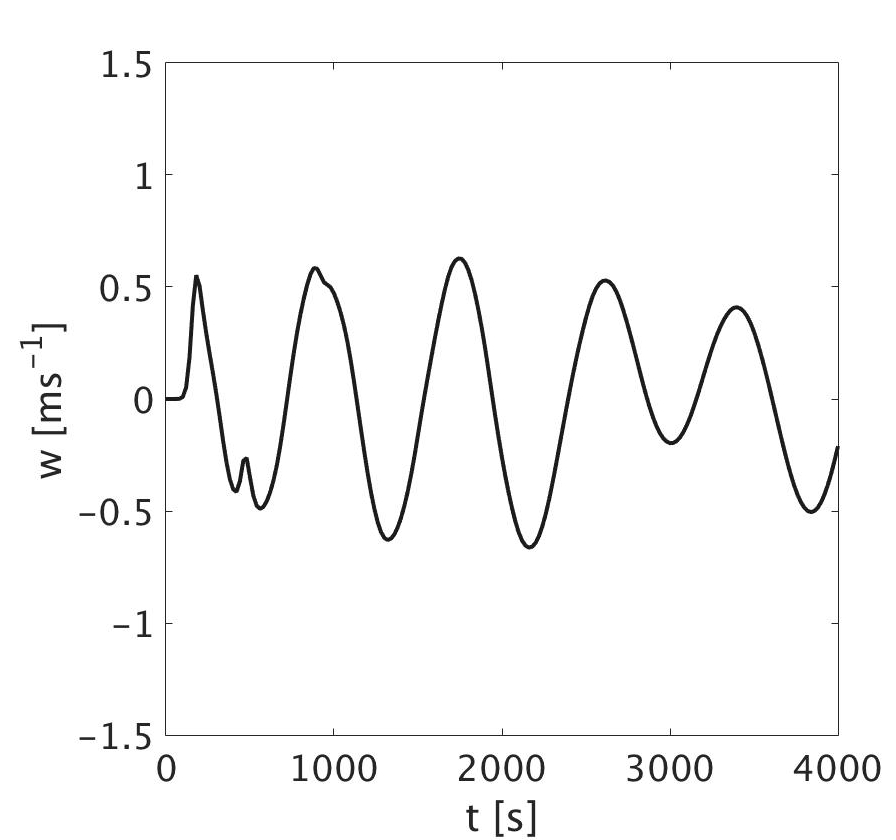}
                \caption{$N = 7.9 \times 10^{-3}$ s$^{-1}$}
                \label{fig:time_series_w_G1_Ri_1}
        \end{subfigure}
\caption{Evolution of vertical velocity at $x = 0$, $z = 6.25$ km for the stable case (a) $N = 2.5 \times 10^{-2}$ s$^{-1}$, (b) $N = 1.25 \times 10^{-2}$ s$^{-1}$, (c) $N = 1.0 \times 10^{-2}$ s$^{-1}$ and (d) $N = 7.9 \times 10^{-3}$ s$^{-1}$. \label{fig:time_series_vertical_velocity}}
\end{figure}

\begin{figure*}
  \centering             
  \begin{subfigure}[b]{0.55\textwidth}
    \centering                      
    \includegraphics[width=19pc,angle=0]{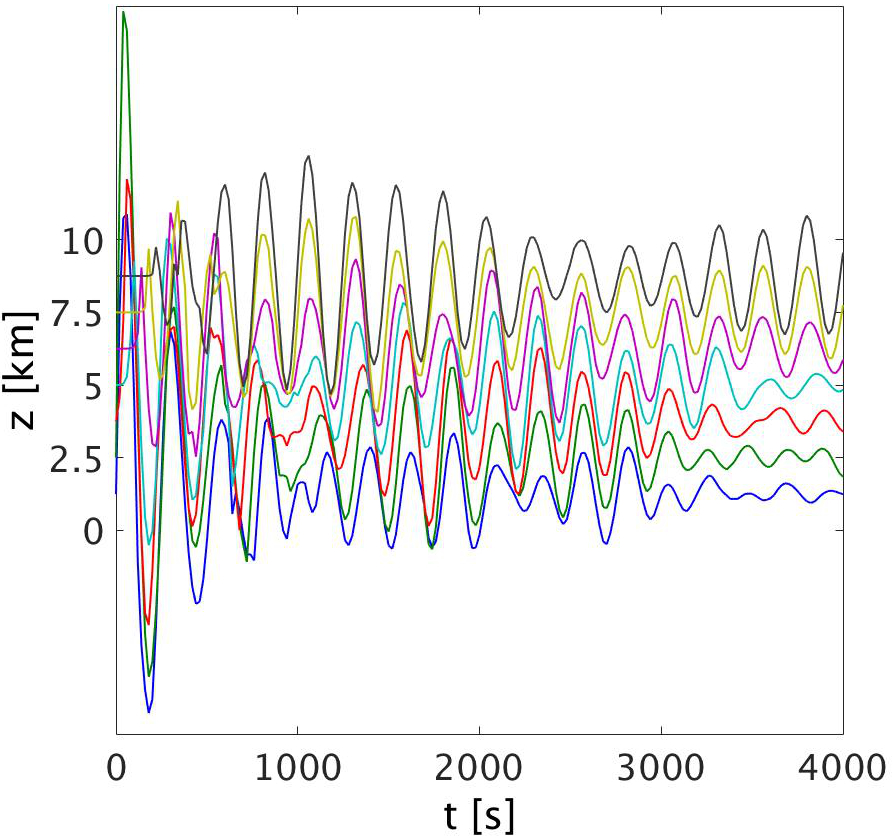}
    \caption{$Ri_b = 1.0, N = 2.5 \times 10^{-2}$ s$^{-1}$}
    \label{fig:time_series_w_G1_Ri1_all1}
  \end{subfigure}
  
  \begin{subfigure}[b]{0.45\textwidth}
    \centering                      
    \includegraphics[width=19pc,angle=0]{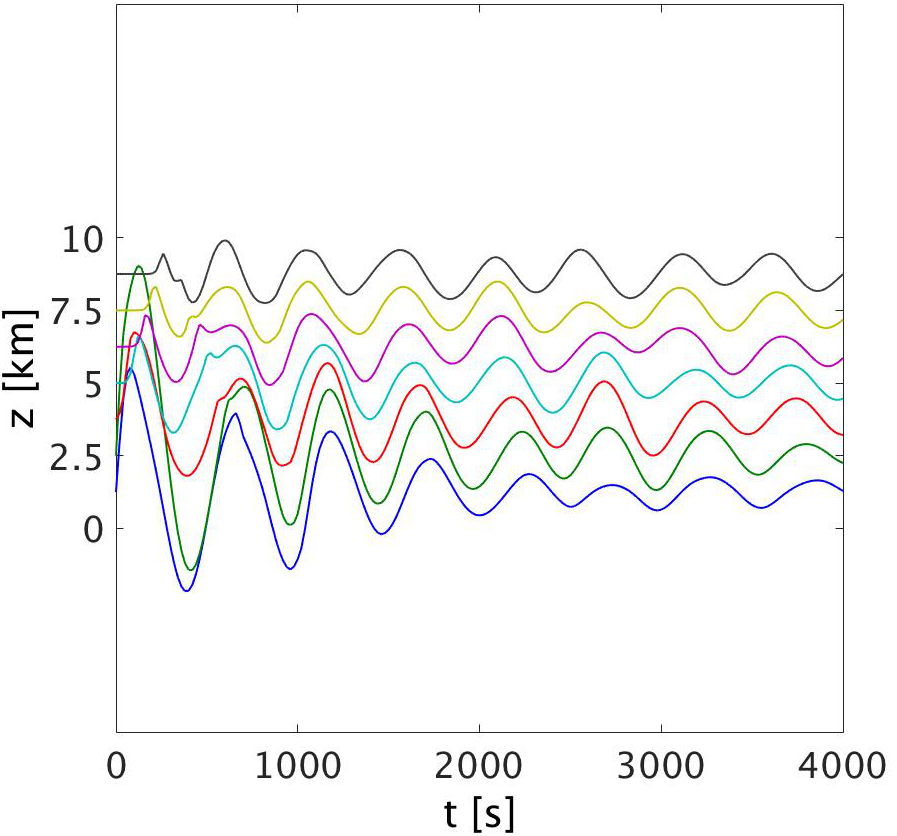}
    \caption{$Ri_b = 0.25,~N = 1.25 \times 10^{-2}$ s$^{-1}$}
    \label{fig:time_series_w_G1_Ri_16_all}
  \end{subfigure}
  ~ %
  \begin{subfigure}[b]{0.45\textwidth}
    \centering                      
    \includegraphics[width=19pc,height=17.8pc,angle=0]{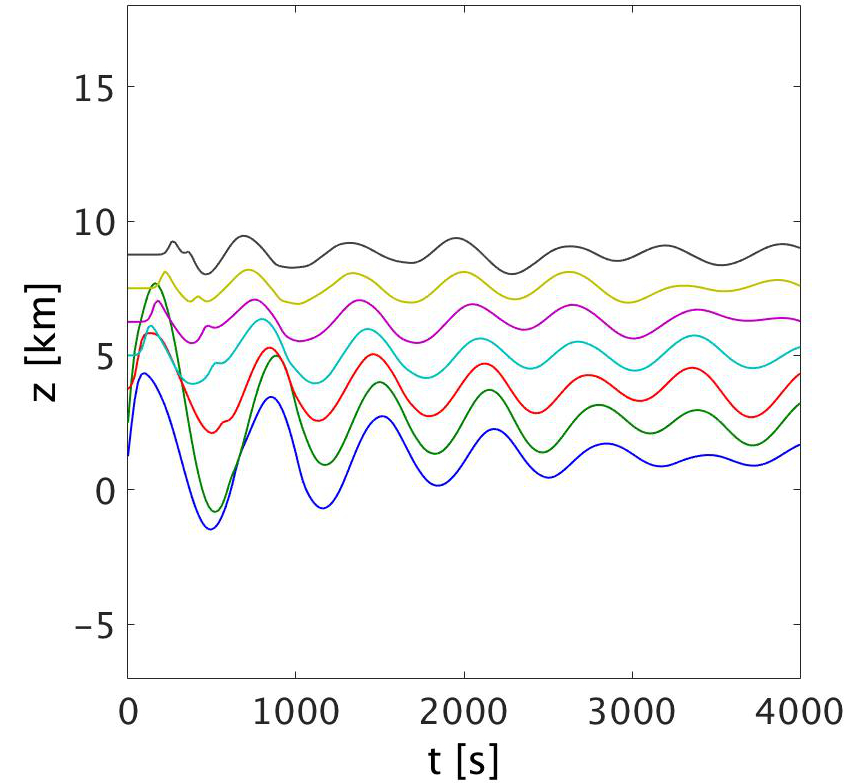}
    \caption{$Ri_b = 0.16,~N = 1.0 \times 10^{-2}$ s$^{-1}$}
    \label{fig:time_series_w_G1_Ri_16_all1}
  \end{subfigure}
  \caption{Time series of vertical velocity at $x = 0$ for various vertical positions $z = 1.25$, $2.5$, $3.75,\, 5.0$, $6.25,\, 7.5,\hbox{ and } 8.75$ km for three values of $N$ \label{fig:time_series_vertical_velocity_1}}
\end{figure*}

Considering a velocity scale of $U \sim 10$ m s$^{-1}$ (based on Table~\ref{tab:extreme_values_ch4_1}) and the buoyancy frequency, $N \sim 1.0 \times 10^{-2}$ s$^{-2}$, one finds a vertical length scale of $L_z = \frac{2 \pi U}{N} \sim 6.25$ km. Thus, we recorded the vertical velocity at a location on the center of the horizontal domain at $z = 6.25$ km for every time step with respect to four values of the buoyancy frequency, $N = 2.5 \times 10^{-2}, 1.25 \times 10^{-2}$, $1.0 \times 10^{-2}, 7.9 \times 10^{-3}$, and the result is shown in Fig~\ref{fig:time_series_vertical_velocity}. For each $N$, the corresponding bulk Richardson numbers $\mathcal Ri_b$, as well as the Froude number $\mathcal Fr$ are presented in Table~\ref{tab:khrare1}. %
For the result in Fig~\ref{fig:time_series_vertical_velocity}, the ratio of the wave frequency to the buoyancy frequency ($\omega/N$) at $z = 6.25$ km are reported in Table \ref{tab:khrare1}. The angle ($\alpha$) between the phase velocity vector and the horizontal direction are also presented in Table \ref{tab:khrare1}, suggesting the dispersion relation: 
\[ \omega = N \cos\alpha. \]
Table \ref{tab:khrare1} also suggests that the wave frequency satisfies $\omega < N$, indicating the maximum possible frequency of internal waves in a stratified fluid is $N$. The angle ($\alpha$) increases as buoyancy frequency decreases. 
To illustrate internal wave propagation, Fig~\ref{fig:time_series_vertical_velocity_1} presents the time series of the vertical velocity corresponding to  seven locations $(0,1.25)$, $(0,2.5)$, $(0,3.75)$, $(0,5.0)$, $(0,6.25)$, and $(0,8.75)$
for three values of $N$. These results show a good agreement of the phenomena simulated by the JFNK method with the corresponding findings reported in the literature~\cite[e.g. see][]{Morton1956}, which means that the wavelet-based JFNK model accurately predicts the penetrative convection of thermals in a stably stratified environment.%

\subsubsection{Conservation of kinetic and potential energy}
The governing equations Eq.~(\ref{eq:ch2_28}-\ref{eq:ch2_30}) leads to the following energy balance laws~\cite[e.g.][]{Winters2009}, where the kinetic and potential energies
\[ E_k = \frac{1}{2} \int_\Omega (u^2+w^2) \mathrm{d}V, \qquad  \mathrm{and} \qquad E_p = \int_\Omega (z_{max} -z) \theta \mathrm{d}V,\]
respectively, satisfy the following energy equations~\cite[see also][e.g. Fig 19 therein]{Carpenter1990}:
\[ \frac{\mathrm{d}E_k}{\mathrm{d}t} =  \int_\Omega w \theta \mathrm{d}V -K_M \epsilon ~,\qquad \epsilon = \int_\Omega |\Delta u|^2+|\Delta w|^2\mathrm{d}V \]
and
\[  \frac{\mathrm{d}E_p}{\mathrm{d}t} = -\int_\Omega w \theta \mathrm{d}V + K_H \frac{\theta_{\max} - \theta_{\min}}{z_{\max} - z_{\min}}.\]
These energy equations quantify the rate of production of $E_p$, the conversion from $E_p$ to $E_k$, and the rate of kinetic energy dissipation, $\epsilon$, thereby making a steady energy balance for the isolated thermal in the neutral environment.

\begin{figure}[t]
  \begin{subfigure}[b]{0.5\textwidth}
    \includegraphics[width=19pc,angle=0]{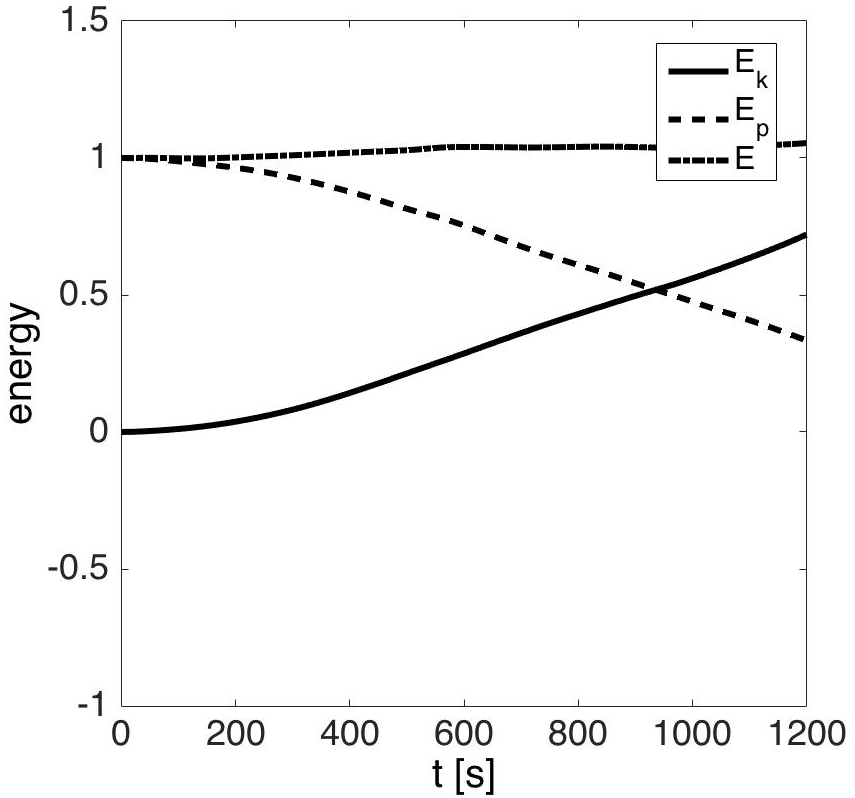}
                \caption{}
                \label{fig:energy_nu1}
        \end{subfigure}
        ~ %
        \begin{subfigure}[b]{0.5\textwidth}
          \includegraphics[width=18.5pc,angle=0]{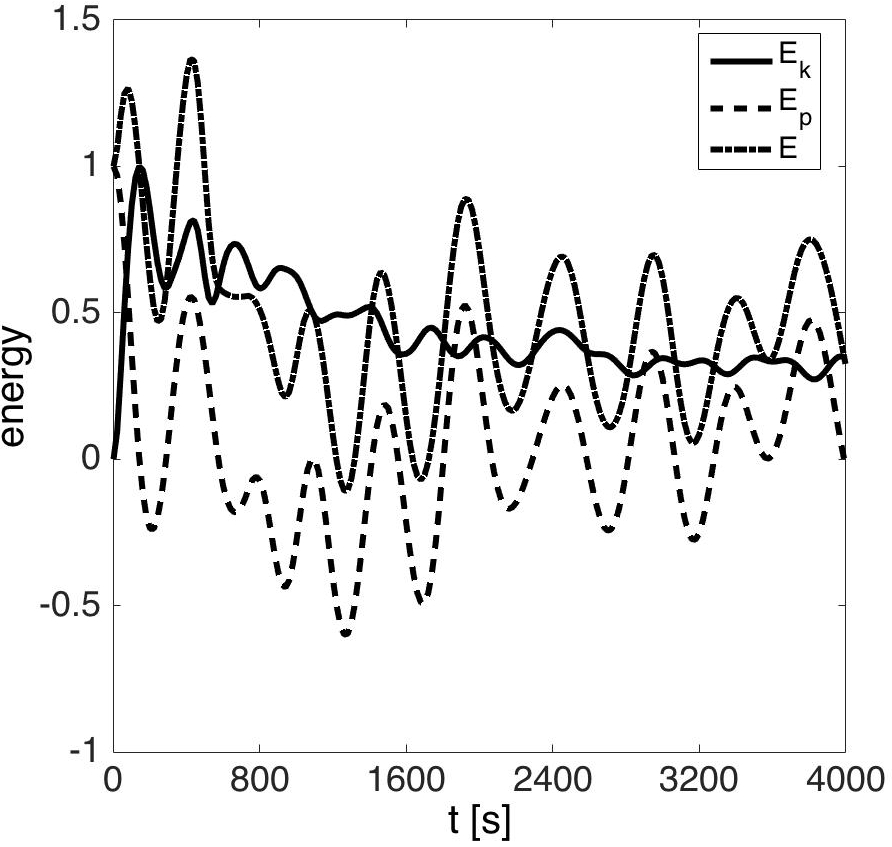}
                \caption{}
                \label{fig:energy_stable}
        \end{subfigure}
              
 \caption{(a) Time evolution of energy for Case A, showing that the total energy is approximately conserved, where potential energy is converted to kinetic energy. Note, $\theta_0 = 300$ K, $K_M = 1$ m$^2$ s$^{-1}$, $t\in[0,1200 s]$. (b) The effect of stratification on the time evolution of energy for Case A, but in the stable environment with $N = 1.25 \times 10^{-2}$ s$^{-1}$ for $t\in[0,4000 s]$.
\label{fig:energy}}
\end{figure}

The time evolution of $E_p$, $E_k$, and $E = E_p + E_k$ for a rising thermal in the neutral environment have been reported in Fig~\ref{fig:energy_nu1}. Clearly, the potential energy $E_p$, decreases with time as a result of the potential energy conversion into kinetic energy $E_k$. Also, the total energy, $E$, remains approximately constant. 
In order to conserve energy, \cite{Carpenter1990} considered the piece-parabolic method for the discretization of convective operators. The present result of the energy conservation depicted in Figure \ref{fig:energy_nu1} has an excellent agreement with the corresponding result reported by~\cite{Carpenter1990}.

However, the time evolution of the potential and kinetic energies feature more complex and oscillating behaviour when the environment is stability stratified. It is because a rising thermal finds itself in an environment with a higher potential temperature, where the buoyancy force pushes it downwards. In Fig~\ref{fig:energy_stable},  the energy curves for $N = 1.25 \times 10^{-2}$ s$^{-1}$ have been displayed. The results clearly indicates the overshooting of thermals beyond their level of buoyancy in the stably stratified environment~\cite[see also][]{Lane2008}. 

\subsection{{Comparison with WRF-LES using} urban heat island circulation}
There is a growing trend of land-surface modification through urbanization because over half the world's population lives in urban areas. {Urban Heat Island~(UHI) is the source of the mesoscale response of the atmosphere to horizontal variations in temperature associated with dry convection~\cite[][]{Grimmond2002}.} Urban Heat Island~(UHI) is a potential atmospheric model to investigate the influence of land-surface modification on the health and welfare of urban residents. UHI simulations help quantify mesoscale circulation triggered by surface heterogeneity of urbanization~\cite[see][]{Zhang2014}.   

\subsubsection{Reference model}
Using the LES mode of the Weather Research and Forecasting (WRF-LES) model, \cite{Zhang2014} investigated the influence of the UHI circulation over an isolated urban area that is homogeneous in $y$-direction, where the atmosphere is dry and the terrain is flat. The WRF-LES model of~\cite{Zhang2014} is similar to the simulation of~\cite{Dubois2009} (hereinafter D~\&~T). Similarly, \cite{Kimura1975} provides a laboratory model of an equivalent UHI circulation. In this section, these three reference models are considered to understand the accuracy of the wavelet-based simulation, where the model domain extends $100$~km horizontally and $2$~km vertically. 

\begin{figure}[htbp]
  \centering
  \begin{tabular}{cc}
    \multicolumn{2}{c}{$(a)$}\\
    \multicolumn{2}{c}{\includegraphics[width=12cm]{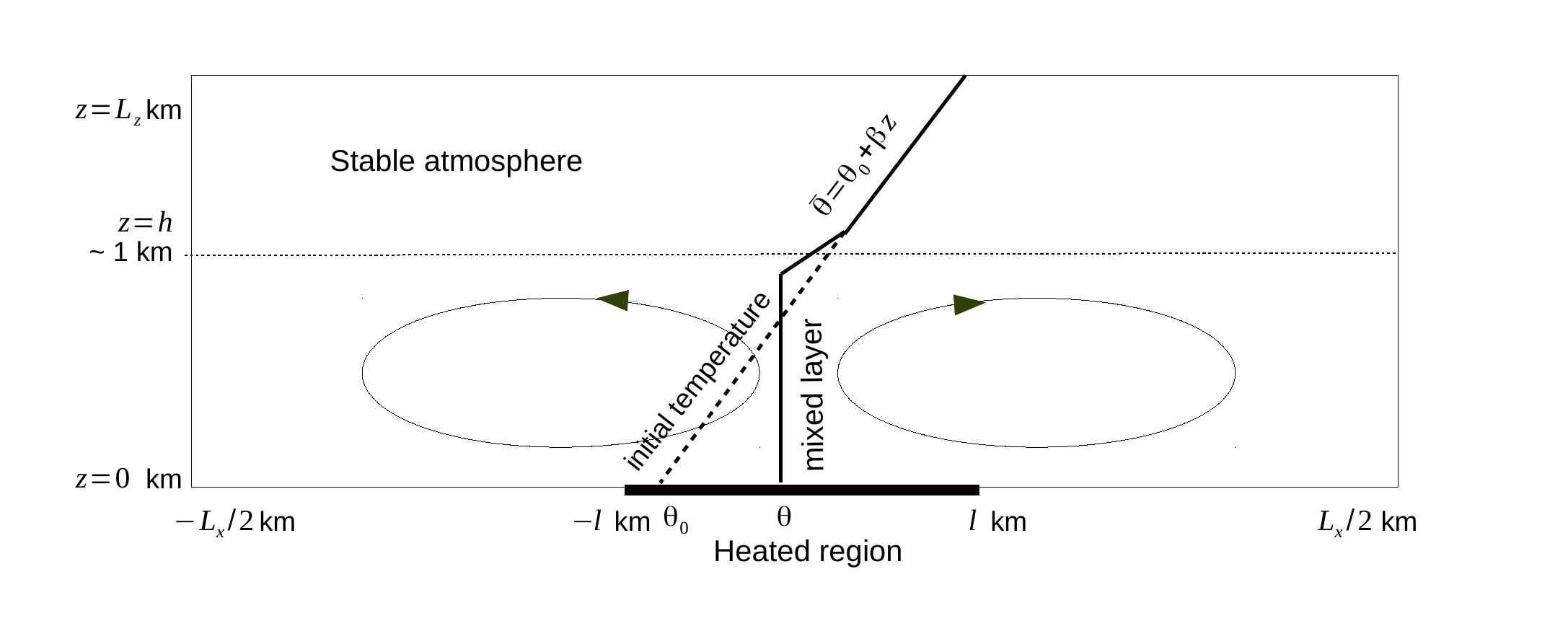}}\\
    $(b)$ & $(c)$\\
    \includegraphics[width=6.75cm]{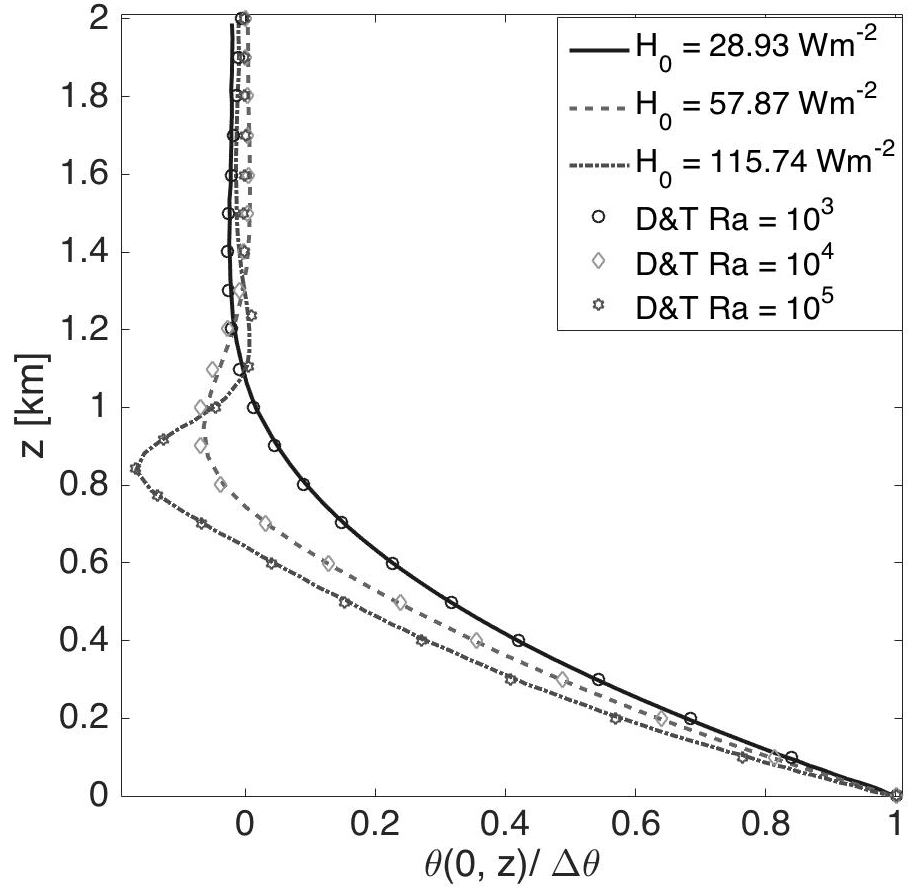}
    &
      \includegraphics[width=7cm]{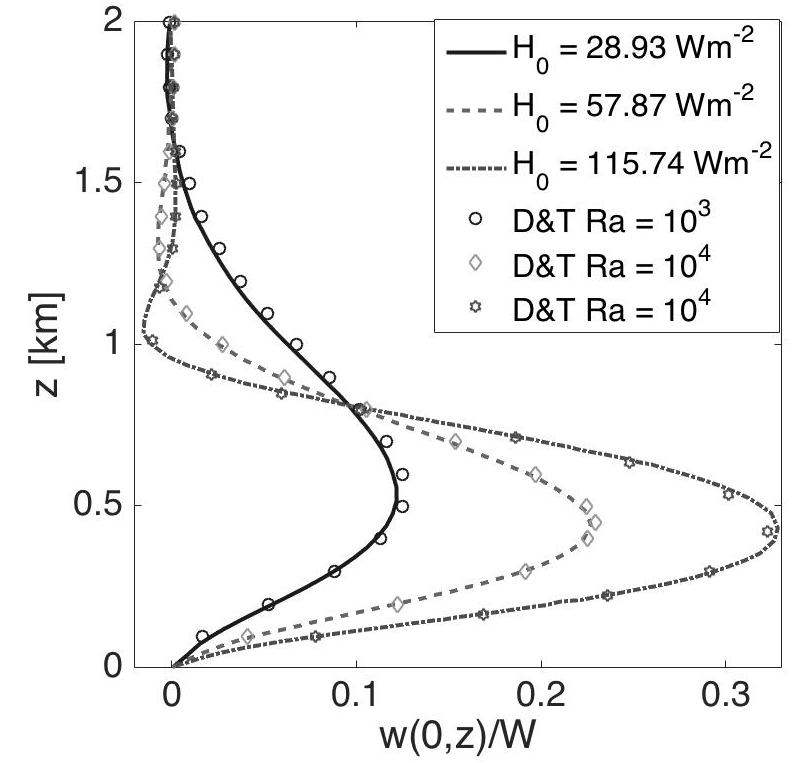}\\
    \multicolumn{2}{c}{$(d)$}\\
    \multicolumn{2}{c}{\includegraphics[width=7cm]{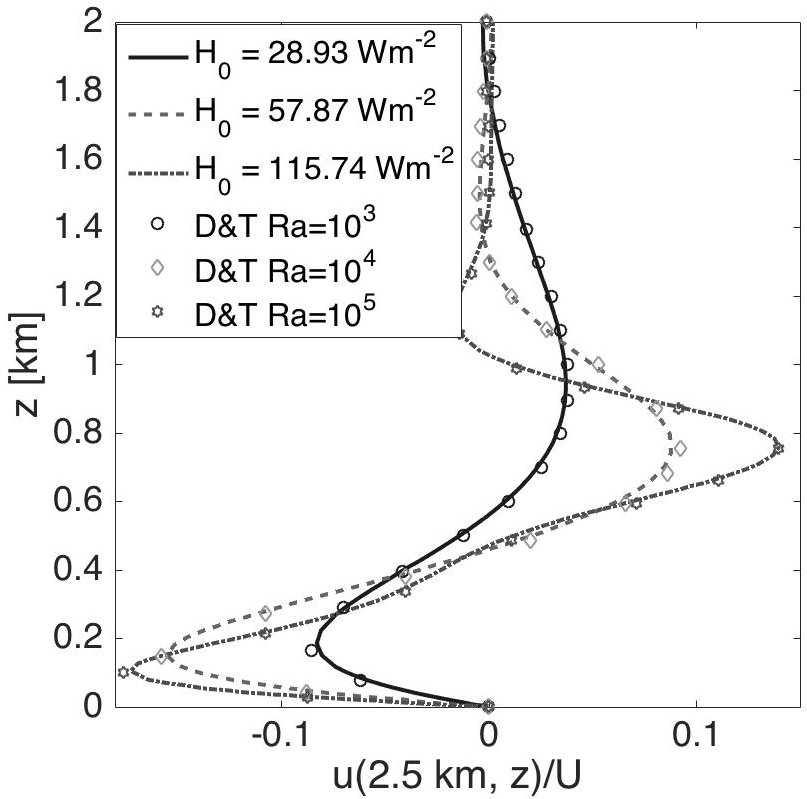}}\\
    
  \end{tabular}
 \caption{(a) A sketch of heat island circulation and the initial profile of the total potential temperature. Profiles of (b) temperature variation $\theta$ at $x = 0$ km, (c) vertical velocity at $x=0$ km and (d) horizontal velocity at $x = 2.5$ km for stationary solution. solid line, dashed line and dashed-dotted line are representing the profiles for $H_0 = 28.93, 57.87$ and $115.74$ W m$^{-2}$, respectively. \label{fig:comparison_dt}}
\end{figure}

Following \cite{Zhang2014} a constant heat flux $(H_{\hbox{surface}}~[\hbox{W m}^{-2}])$ and $u = w = 0$ are boundary conditions at the surface, $z = 0$, such that 
$$
H_{\hbox{surface}}-H_{\hbox{rural}} = \left\{
  \begin{array}{ll}
    0 & \hbox{ if } x < -l \hbox{ or } x > l\\
    H_0\tanh\left(\frac{x+1}{\xi}\right)-H_0\tanh\left(\frac{x-1}{\xi}\right) & \hbox{ if }  -l \le x \le l.
  \end{array}
\right.,
$$
where $H_{\hbox{rural}}$ is the surface heat flux over the rural area and $H_0$ is the surface heat flux over the urban area (see Fig~\ref{fig:comparison_dt}$a$).

Values of $H_0 = 28.93$, $57.87,\,115.74,\,231.48,\,462.96$, and $926.92$~[W~m$^{-2}$] were tested. These six values of $H_0$ correspond to six values of Rayleigh numbers $\mathcal Ra = 10^3$, $10^4,\,10^5,\,10^6,\,10^7,\,\hbox{ and } 10^8$.

\begin{table*}[t]
\caption {The effect of surface heat flux variation on the potential temperature $\theta_{\min}$, horizontal velocity $u_{\max}$, vertical velocity $w_{\max}$, and vorticity $\omega_{\max}$ has been predicted accurately by the present model in comparison to the results of \cite{Dubois2009}, {\em i.e.} D\&T } 
\label{tab:comparison_dubois}
\begin{center}
\begin{tabular}{crrrrrr}
\hline\hline
$H_0$ & \multicolumn{2}{ c }{$28.93$ W m$^{-2}$}& \multicolumn{2}{ c }{ $57.87$ W m$^{-2}$} &\multicolumn{2}{ c }{ $115.74$ W m$^{-2}$}\\
\hline
&Present&D \& T&Present&D \& T&Present&D \& T\\
\hline
$\theta_{\min}$ & -0.023537  & -0.024823& -0.064457 &-0.071289 &  -0.167264  & -0.166316\\ 
$u_{\max}$ & 0.118872 &0.118887 &  0.176103 &0.174844 & 0.179622 &0.179054\\
$w_{\min}$ &  -0.030134 & -0.030470& -0.037337  & -0.039291&   -0.085519 &-0.079265\\
$\omega_{\max}$ & 1.957423 & 2.06900 & 3.659917 & 3.951325 & 5.345340 & 5.921375 \\
\hline
\end{tabular}
\end{center}
\end{table*}

\subsection{Flow evolution at relatively smaller values of the surface heat flux}\label{comparison_dubois}
Fig~\ref{fig:comparison_dt} demonstrates the vertical profiles of the vertical velocity and the potential temperature computed at the centre of the heat island for $H_0 = 28.93, 57.87$ and $115.74$~Wm$^{-2}$. In  dimensionless variables, these plots are found in a very good agreement with the corresponding plots of~\cite{Dubois2009}. The temperature, vertical velocity, and horizontal velocity decay rapidly with respect to the elevation $z$. 
In Figure \ref{fig:comparison_dt}b and \ref{fig:comparison_dt}c, it is clear that the mixed layer height appears between $z = 1.2$~km and $1.5$~km, and this height is reduced if $H_0$ increases.
The values of the horizontal velocity $u$, the vertical velocity $w$, the potential temperature $\theta$, and the vorticity $\omega_y = \partial u/\partial z - \partial w/\partial x$ were compared between the results of the present model and that of~\cite{Dubois2009}. The results presented in Table~\ref{tab:comparison_dubois} indicates that the accuracy of the wavelet-based JFNK model is within $5\%$ to $10\%$ of the results of D~\&~T for the test case of UHI.

\subsection{Experimental investigation for penetrative convection}
\begin{figure}[t]
       \noindent\includegraphics[width=\textwidth]{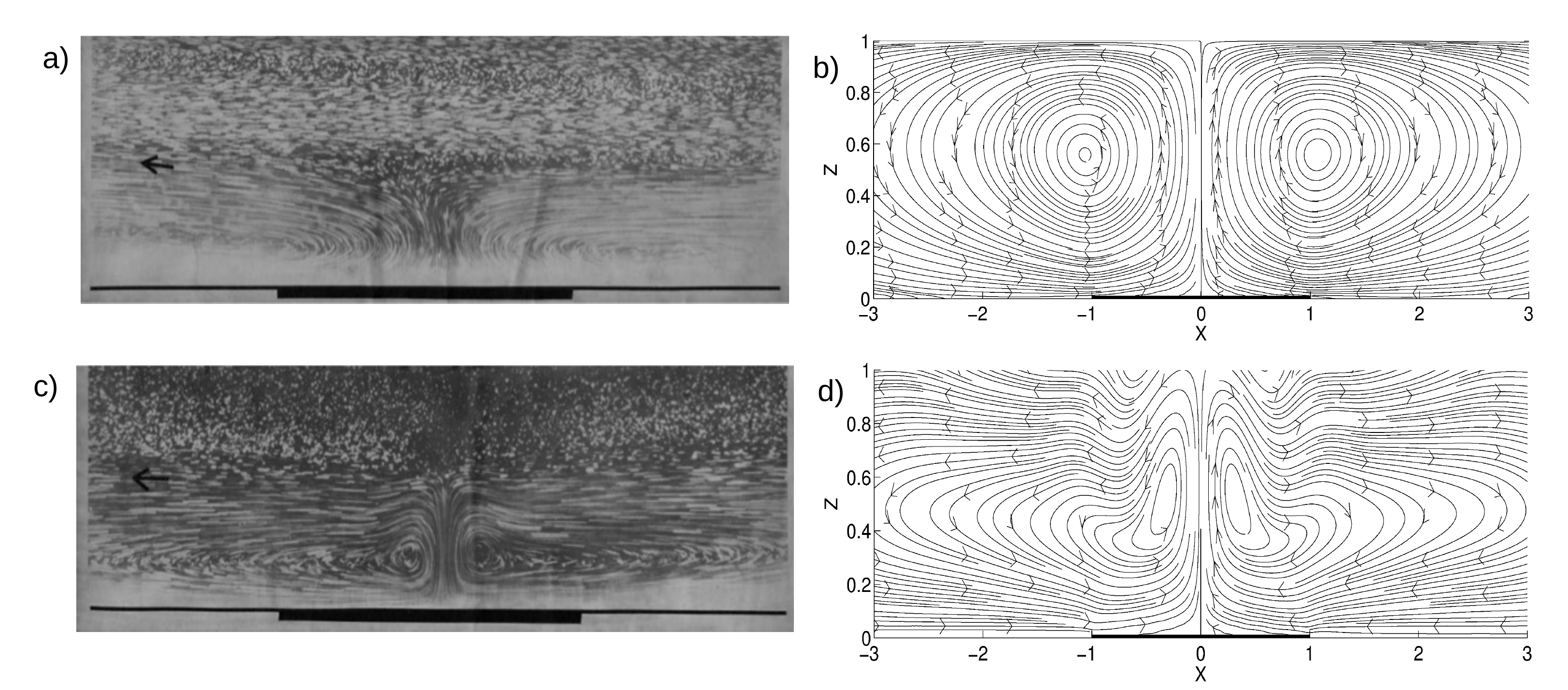}
 \caption{(Left) Two types of flow regimes observed in a laboratory experiment~\cite[][]{Kimura1975}. (a) Center of the circulation is located near the edge of the heating element; (c) the center of the circulation is near the center of the heating element. (Right) Numerically experimented results: (b) $H_0 = 28.93$ W m$^{-2}$, $\beta = 1$ K km$^{-1}$ (d) $H_0 = 115.74$ W m$^{-2}$, $\beta = 10$ K km$^{-1}$. Thick black horizontal line represents the heating element. Note, the center of the circulation approximately corresponds to each other in $(a-b)$ and $(c-d)$. The thick arrow on the left panel is about $z=1$ on the right panel. The domain is made dimensionless by half the length of heating element. \label{fig:vorticitycontour_uhi_ex}}
\end{figure}

It was observed in the experimental study of \cite{Kimura1975}
that the centre of the heat island circulation is located near the edges of the heat island  when the surface heat flux is relatively weak, and the up-draft prevails all over the urban area (e.g Fig. \ref{fig:vorticitycontour_uhi_ex}a). On the other hand, a strong narrow up-draft is concentrated above the centre of the island, when the heat flux is relatively strong (e.g. Fig. \ref{fig:vorticitycontour_uhi_ex}c). 

The simulation results in Fig~\ref{fig:vorticitycontour_uhi_ex}b and \ref{fig:vorticitycontour_uhi_ex}d are in a very good agreement with the corresponding experimental results.  It was found that if $H_0$ increases, the center of the circulation moves toward the center of the urban region.
In  Fig. \ref{fig:vorticitycontour_uhi_ex}b, the surface heat flux is $28.93$ W m$^{-2}$, where the center of the circulation is away from the center of the heat island, and in Fig. \ref{fig:vorticitycontour_uhi_ex}d, the surface heat flux is $115.74$ W m$^{-2}$, where the center of the circulation is at the center of the heat island. 

\subsection{A comparison with field measurement}

To provide a primary assessment of the model for predicting the structure of a convective boundary layer~(CBL), an idealized dry case is studied, which is driven solely by a surface heat flux. The result is analyzed with respect to the field measurements of the day-33 Wangara experiment~\cite[e.g.][]{Moeng84}, where the temporal evolution of the mixed layer is studied. In a similar study, \cite{Moeng84} considered LES to reproduce Wangara data when the numerical model incorporates a prameterization of the moisture field, radiation effect, Coriolis effect, and surface roughness in addition to having the wind shear consistent with the field measurements. \cite{Mukherjee2016} provides an idealized numerical study of the CBL using a turbulence-resolving LES in the domain $3.6\times 3.6\times 1.9$~km$^3$.

In comparison to the reference work mentioned above, the present study considers a simplified case in a two-dimensional domain. Here, the horizontal length of $100$~km is divided into $512$ segments ($\Delta x\approx 195$~m), and the vertical length of $2$~km is divided into $256$ segments ($\Delta z\approx 8$~m).
Horizontally $\langle\cdot\rangle$ and temporally $\overline{(\cdot)}$ averaged vertical profile of the total potential temperature, {\em i.e.} $\langle\bar\theta\rangle = \langle\overline{\theta_0 + \beta z + \theta}\rangle$, for $H_0 = 462.96$~W~m$^{-2}$ and $H_0 = 925.92$~W~m$^{-2}$ are analyzed. We observed that the estimated mixed layer height, for $H_0 = 462.96$ W m$^{-2}$ and $H_0 = 925.92$ W m$^{-2}$, are about $0.9$ km and $0.8$ km, respectively, and the inversion layer appears about at $0.9 - 1.0$ km and $0.8 - 0.9$ km from the ground, respectively.  

\begin{figure}
  \centering
  \begin{tabular}{c}
    %
    %
    \includegraphics[width=8cm,height=10cm]{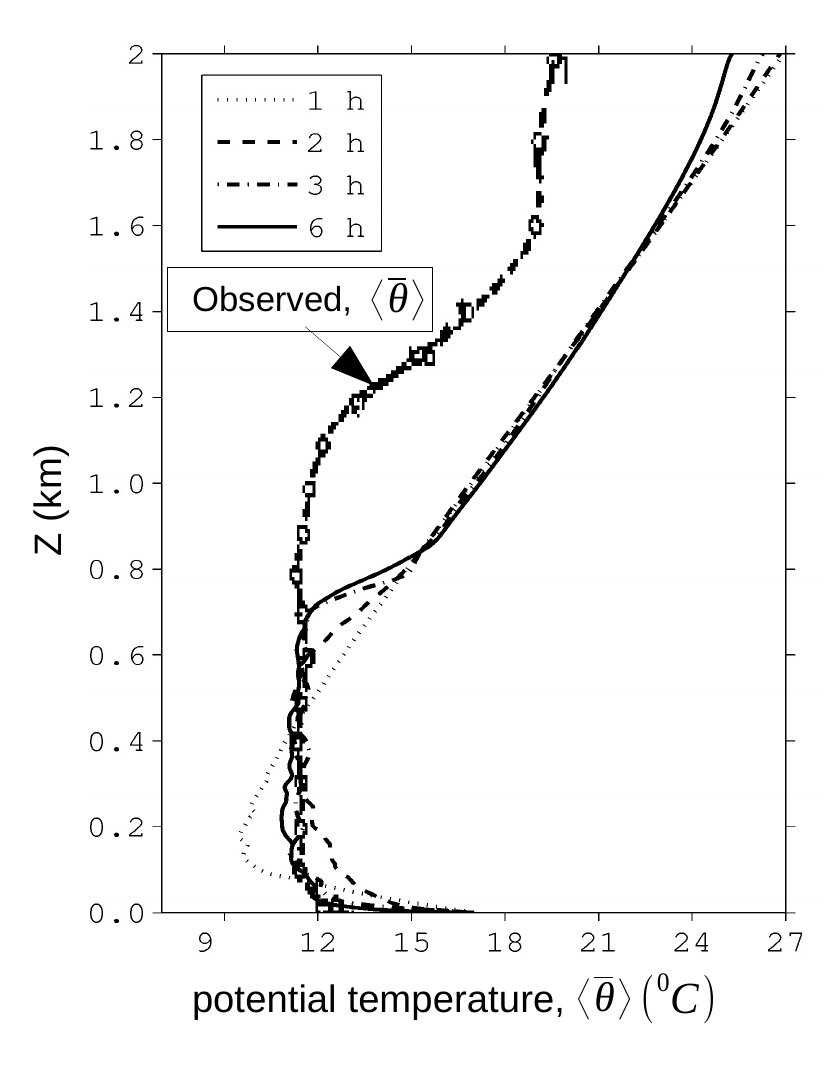}
    \\
  \end{tabular}
 \caption{ The vertical profile of the average potential temperature $\langle\bar\theta\rangle$. The temporal average is obtained in the time interval of $t\in[0$ h, $6$ h$]$. The prediction of the JFNK model at $H_0=925.92$ W m$^{-2}$ is compared with the field measurement (by digitally extracting the data from Fig~3b of~\cite{Moeng84}), which represents the vertical profile of $\langle\bar\theta\rangle$ at $12$-th hour of day $33$. The hourly evolution of the vertical temperature profile is also shown for clarity. Note for the unit conversion to $^{\hbox{\tiny o}}$C,  which aims for being consistent with~\cite{Moeng84}. \label{fig:ts_av_theta_x_0}}
\end{figure}

In Figure \ref{fig:ts_av_theta_x_0}, we compare the vertical profile of $\langle\bar\theta\rangle(z)$ of the present simulation with the similar profile observed in the Wangara day-33 experiment~\cite[{\em e.g.},][]{Moeng84}. The displayed data corresponds to the surface heat flux of $H_0 = 925.92$ W m$^{-2}$. We can see the development of a well-mixed layer within $2$~h. The mixed layer in the turbulent region is capped with the inversion layer approximately at $z\sim 800$~m. At the bottom of the boundary layer, the unstable surface layer appears at $\sim 50$~m. The mixed layer height did not fully agree between the simulation and the measurement because the idealized simulation ignored the realistic meteorological features, such as wind-shear, Coriolis effect etc. However, the vertical structure of the daytime boundary layer has been predicted with a good quality. 

\section{Summary and future developments}\label{sec:sfd}

A symmetry preserving JFNK method for accurate simulations of nonlinearly coupled atmospheric physics has been illustrated in this article. The method has been applied to simulate atmospheric convection using the nonhydrostatic atmospheric model equations. Unlike linearized methods commonly adopted in atmospheric flow solvers, the JFNK method emphasizes that nonlinear coupling of physics to be modelled without any compromise. In atmospheric  flow simulations, capturing the tight nonlinear coupling may lead to a next generation atmospheric model to correctly forecast  weather events. In this work we have shown that preserving the skew-symmetry of the convective operator by the wavelet method brings two modelling benefits. First, short-wavelength modes are properly cascaded toward the subgrid scale dissipation mechanism. This is done by ensuring the role of the convective operator it would play on the physics of the flow. Second, the subgrid-scale modes are properly transferred to the subgrid model for being dissipated at a rate offered by the subgrid model.

In mathematical terms, an interplay between the skew-symmetric convection and symmetric, negative-definite diffusion leads to small-scale motion in a turbulent flow. With this hypothesis of energy cascade in mind, we have combined the JFNK method with a symmetry-preserving descretization that is based on the wavelet method. This article presents the efficiency and reliability of the JFNK methodology for numerical simulation of nonhydrostatic atmospheric flows in the context of penetrative convection. We have chosen to simulate idealized dry convection for the presentation of the JFNK method because the evolution of plumes and thermals offers much insight into the more complicated dynamics of atmospheric motion. Because of the tightly nonlinear coupling of atmospheric motions, present authors envision to perform simulations in such a manner that the discretized operators preserve the same symmetry properties and the same nonlinear coupling as the underlying differential operators. The main question becomes whether the symmetry-preserving nonlinearly coupled discretization is appropriate for atmospheric simulations since the atmospheric modelling community has accepted the discretization of the skew-symmetric convective opterator to a positive-definite convective operator in their publicly available codes~\cite[see][and the refs therein]{Klemp2007,Piotr2014,Smolarkiewicz2017}. This question has been addressed by simulations for which conservation of energy is highly desired to correctly forecast intensification of particular weather events~\cite[see][]{Carpenter1990,Klemp2007}.

The tight nonlinear coupling of physics offered by the JFNK solver will bring full benefit to atmospheric simulations if a scale-adaptive subgrid model is considered~\cite[see][]{Alam2018}. Transition of scales -- often labelled as the `gray-zone' -- is currently one of the most challenging problems in the field of meteorology~\cite[][]{Wyngaard2004,Kurowski2018}. The findings of this article encourages to further test the JFNK method using a more appropriate scale-adaptive subgrid model that adapts the cut-off scale dynamically as the characteristic scale exhibits transitions. For example, a balance between the local production and the dissipation of turbulence occurs in the surface-layer at much smaller scales than the characteristic scale of eddies in the outer layer. In the future research, we plan on developing wavelet-based preconditioners based on lifting schemes, which would discretize the differential operators on to a hierarchy of `details' wavelet space $\mathcal V^{j+1}\backslash\mathcal V^j$ instead of the approximation space $\mathcal V^j$ considered in the present work. 

\section*{Acknowledgements}
JA acknowledges financial support from Natural Science and Engineering Research Council (NSERC) in the form of a Discovery Grant. The article was benefited from comments from two anonymous reviewers. This research was enabled in part with support provided by SHARCNET (www.sharcnet.ca) and Compute Canada (www.computecanada.ca).

\bibliographystyle{model1-num-names}
\bibliography{alamj2020} 

\begin{thebibliography}{38}
\expandafter\ifx\csname natexlab\endcsname\relax\def\natexlab#1{#1}\fi
\providecommand{\bibinfo}[2]{#2}
\ifx\xfnm\relax \def\xfnm[#1]{\unskip,\space#1}\fi
\bibitem[{Ferziger and Peric(1997)}]{Ferziger97}
\bibinfo{author}{J.~Ferziger}, \bibinfo{author}{M.~Peric},
  \bibinfo{title}{Computational Methods for Fluid Dynamics},
  \bibinfo{publisher}{Springer}, \bibinfo{address}{New York},
  \bibinfo{year}{1997}.
\bibitem[{Pielke(2002)}]{Pielke2002}
\bibinfo{author}{R.~A. Pielke}, \bibinfo{title}{Mesoscale Meteorological
  Modeling}, \bibinfo{publisher}{Academic Press}, \bibinfo{edition}{2nd}
  edition, \bibinfo{year}{2002}.
\bibitem[{Pletcher et~al.(2013)Pletcher, Tannehill, and
  Anderson}]{Tannehill2013}
\bibinfo{author}{R.~H. Pletcher}, \bibinfo{author}{J.~C. Tannehill},
  \bibinfo{author}{D.~A. Anderson}, \bibinfo{title}{Computational Fluid
  Mechanics and Heat transfer}, \bibinfo{publisher}{Taylor \& Francis group},
  \bibinfo{edition}{third} edition, \bibinfo{year}{2013}.
\bibitem[{Wicker and Skamarock(1998)}]{Wicker1998}
\bibinfo{author}{L.~J. Wicker}, \bibinfo{author}{W.~C. Skamarock},
\newblock \bibinfo{title}{A time-splitting scheme for the elastic equations
  incorporating second-order runge-kutta time differencing.},
\newblock \bibinfo{journal}{Mon. Wea. Rev.} \bibinfo{volume}{126}
  (\bibinfo{year}{1998}) \bibinfo{pages}{1992--1999}.
\bibitem[{Skamarock et~al.(1997)Skamarock, Smolarkiewicz, and
  Klemp}]{Skamarock97}
\bibinfo{author}{W.~C. Skamarock}, \bibinfo{author}{P.~K. Smolarkiewicz},
  \bibinfo{author}{J.~B. Klemp},
\newblock \bibinfo{title}{Preconditioned conjugate-residual solvers for
  helmholtz equations in nonhydrostatic models},
\newblock \bibinfo{journal}{Monthly Weather Review} \bibinfo{volume}{125}
  (\bibinfo{year}{1997}) \bibinfo{pages}{587--599}.
\bibitem[{Verstappen and Veldman(2003)}]{Verstappen2003}
\bibinfo{author}{R.~Verstappen}, \bibinfo{author}{A.~Veldman},
\newblock \bibinfo{title}{Symmetry-preserving discretization of turbulent
  flow},
\newblock \bibinfo{journal}{Journal of Computational Physics}
  \bibinfo{volume}{187} (\bibinfo{year}{2003}) \bibinfo{pages}{343 -- 368}.
\bibitem[{Steppeler et~al.(2003)Steppeler, Hess, Sch{\"a}ttler, and
  Bonaventura}]{Steppeler2003}
\bibinfo{author}{J.~Steppeler}, \bibinfo{author}{R.~Hess},
  \bibinfo{author}{U.~Sch{\"a}ttler}, \bibinfo{author}{L.~Bonaventura},
\newblock \bibinfo{title}{Review of numerical methods for nonhydrostatic
  weather prediction models},
\newblock \bibinfo{journal}{Meteorology and Atmospheric Physics}
  \bibinfo{volume}{82} (\bibinfo{year}{2003}) \bibinfo{pages}{287--301}.
\bibitem[{Klemp et~al.(2007)Klemp, Skamarock, and Dudhia}]{Klemp2007}
\bibinfo{author}{J.~B. Klemp}, \bibinfo{author}{W.~C. Skamarock},
  \bibinfo{author}{J.~Dudhia},
\newblock \bibinfo{title}{Conservative split-explicit time integration methods
  for the compressible nonhydrostatic equations},
\newblock \bibinfo{journal}{Monthly Weather Review} \bibinfo{volume}{135}
  (\bibinfo{year}{2007}) \bibinfo{pages}{2897--2913}.
\bibitem[{Bryan and Fritsch(2002)}]{Bryan2002}
\bibinfo{author}{G.~H. Bryan}, \bibinfo{author}{J.~M. Fritsch},
\newblock \bibinfo{title}{A benchmark simulation for moist nonhydrostatic
  numerical model},
\newblock \bibinfo{journal}{Mon. Wea. Rev.} \bibinfo{volume}{130}
  (\bibinfo{year}{2002}).
\bibitem[{Knoll and Keyes(2004)}]{Knoll2004}
\bibinfo{author}{D.~A. Knoll}, \bibinfo{author}{D.~E. Keyes},
\newblock \bibinfo{title}{Jacobian-free newton-krylov methods: a survey of
  approaches and applications},
\newblock \bibinfo{journal}{J. Comput. Phys.} \bibinfo{volume}{193}
  (\bibinfo{year}{2004}) \bibinfo{pages}{357--397}.
\bibitem[{Chisholm and Zingg(2009)}]{Zingg2009}
\bibinfo{author}{T.~T. Chisholm}, \bibinfo{author}{D.~W. Zingg},
\newblock \bibinfo{title}{A jacobian-free newton–krylov algorithm for
  compressible turbulent fluid flows},
\newblock \bibinfo{journal}{Journal of Computational Physics}
  \bibinfo{volume}{228} (\bibinfo{year}{2009}) \bibinfo{pages}{3490 -- 3507}.
\bibitem[{Carpenter et~al.(1990)Carpenter, Droegemeier, Woodward, and
  Hane}]{Carpenter1990}
\bibinfo{author}{R.~L.~J. Carpenter}, \bibinfo{author}{K.~K. Droegemeier},
  \bibinfo{author}{P.~R. Woodward}, \bibinfo{author}{C.~E. Hane},
\newblock \bibinfo{title}{Application of the piecewise parabolic method(ppm) to
  meteorological modelling.},
\newblock \bibinfo{journal}{Mon. Wea. Rev.} \bibinfo{volume}{118}
  (\bibinfo{year}{1990}) \bibinfo{pages}{586--612}.
\bibitem[{Lane(2008)}]{Lane2008}
\bibinfo{author}{T.~P. Lane},
\newblock \bibinfo{title}{The vortical response to penetrative convection and
  the associated gravity-wave generation},
\newblock \bibinfo{journal}{Atmos. Sci. Lett.} \bibinfo{volume}{9}
  (\bibinfo{year}{2008}) \bibinfo{pages}{103--110}.
\bibitem[{Mousseau et~al.(2002)Mousseau, Knoll, and Reisner}]{Reisner2002}
\bibinfo{author}{V.~A. Mousseau}, \bibinfo{author}{D.~A. Knoll},
  \bibinfo{author}{J.~M. Reisner},
\newblock \bibinfo{title}{An implicit nonlinearly consistent method for the
  two-dimensional shallow-water equations with coriolis force},
\newblock \bibinfo{journal}{Monthly Weather Review} \bibinfo{volume}{130}
  (\bibinfo{year}{2002}) \bibinfo{pages}{2611--2625}.
\bibitem[{Alam and Islam(2015)}]{Alam2015}
\bibinfo{author}{J.~Alam}, \bibinfo{author}{M.~R. Islam},
\newblock \bibinfo{title}{A multiscale eddy simulation methodology for the
  atmospheric ekman boundary layer},
\newblock \bibinfo{journal}{Geophysical \& Astrophysical Fluid Dynamics}
  \bibinfo{volume}{109} (\bibinfo{year}{2015}) \bibinfo{pages}{1--20}.
\bibitem[{Dubois and Touzani(2009)}]{Dubois2009}
\bibinfo{author}{T.~Dubois}, \bibinfo{author}{R.~Touzani},
\newblock \bibinfo{title}{A numerical study of heat island flows: Stationary
  solutions},
\newblock \bibinfo{journal}{International Journal for Numerical Methods in
  Fluids} \bibinfo{volume}{59} (\bibinfo{year}{2009})
  \bibinfo{pages}{631--655}.
\bibitem[{LeVeque(1990)}]{Randall90}
\bibinfo{author}{R.~J. LeVeque}, \bibinfo{title}{Numerical Methods for
  Conservation Laws}, \bibinfo{publisher}{Birkhauser Verlag},
  \bibinfo{address}{Boston}, \bibinfo{year}{1990}.
\bibitem[{Deslauriers and Dubuc(1989)}]{Deslauriers1989}
\bibinfo{author}{G.~Deslauriers}, \bibinfo{author}{S.~Dubuc},
\newblock \bibinfo{title}{Symmetric iterative interpolation process.},
\newblock \bibinfo{journal}{Constructive Approximation} \bibinfo{volume}{5}
  (\bibinfo{year}{1989}) \bibinfo{pages}{49--68}.
\bibitem[{Mallat(2009)}]{Mallat2009}
\bibinfo{author}{S.~Mallat}, \bibinfo{title}{A wavelet tour of signal
  processing}, \bibinfo{publisher}{Academic Press}, \bibinfo{year}{2009}.
\bibitem[{Alam et~al.(2014)Alam, Walsh, Alamgir~Hossain, and Rose}]{Alam2014}
\bibinfo{author}{J.~M. Alam}, \bibinfo{author}{R.~P. Walsh},
  \bibinfo{author}{M.~Alamgir~Hossain}, \bibinfo{author}{A.~M. Rose},
\newblock \bibinfo{title}{A computational methodology for two-dimensional fluid
  flows},
\newblock \bibinfo{journal}{International Journal for Numerical Methods in
  Fluids} \bibinfo{volume}{75} (\bibinfo{year}{2014})
  \bibinfo{pages}{835--859}.
\bibitem[{Deardorff(1970)}]{dear70}
\bibinfo{author}{J.~W. Deardorff},
\newblock \bibinfo{title}{A three-dimensional numerical investigation of
  idealized planetary boundary layer},
\newblock \bibinfo{journal}{Geophys. Fluid Dyn.} \bibinfo{volume}{1}
  (\bibinfo{year}{1970}) \bibinfo{pages}{377--410}.
\bibitem[{Deardorff(1980)}]{Deardorff80}
\bibinfo{author}{J.~W. Deardorff},
\newblock \bibinfo{title}{Stratocumulus-capped mixed layers derived from a
  three-dimensional model},
\newblock \bibinfo{journal}{Boundary-Layer Meteorology} \bibinfo{volume}{18}
  (\bibinfo{year}{1980}) \bibinfo{pages}{495--527}.
\bibitem[{Moeng and Sullivan(2015)}]{Moeng2015}
\bibinfo{author}{C.-H. Moeng}, \bibinfo{author}{P.~Sullivan},
  \bibinfo{title}{Encyclopedia of Atmospheric Sciences, 2nd Edition},
  volume~\bibinfo{volume}{4}, \bibinfo{publisher}{Elsevier Ltd, Academic
  Press}, pp. \bibinfo{pages}{232--240}.
\bibitem[{Bartello and Tobias(2013)}]{Bartello2013}
\bibinfo{author}{P.~Bartello}, \bibinfo{author}{S.~M. Tobias},
\newblock \bibinfo{title}{Sensitivity of stratified turbulence to the buoyancy
  reynolds number},
\newblock \bibinfo{journal}{Journal of Fluid Mechanics} \bibinfo{volume}{725}
  (\bibinfo{year}{2013}) \bibinfo{pages}{1–22}.
\bibitem[{Saad and Schultz(1986)}]{Saad1986}
\bibinfo{author}{Y.~Saad}, \bibinfo{author}{M.~H. Schultz},
\newblock \bibinfo{title}{Gmres: a generalized minimal residual algorithm for
  solving nonsymmetric linear systems},
\newblock \bibinfo{journal}{SIAM J. Sci. Stat. Comput.} \bibinfo{volume}{7}
  (\bibinfo{year}{1986}) \bibinfo{pages}{856--869}.
\bibitem[{Lin(2007)}]{Lin2007}
\bibinfo{author}{Y.-L. Lin}, \bibinfo{title}{Mesocale Dynamics},
  \bibinfo{publisher}{Cambridge University Press}, \bibinfo{year}{2007}.
\bibitem[{Morton et~al.(1956)Morton, Taylor, and Turner}]{Morton1956}
\bibinfo{author}{B.~R. Morton}, \bibinfo{author}{G.~Taylor},
  \bibinfo{author}{J.~S. Turner},
\newblock \bibinfo{title}{Turbulent and gravitational convection from
  maintained and instaneous sources},
\newblock \bibinfo{journal}{Proceedings of the Royal Society of London}
  \bibinfo{volume}{A234} (\bibinfo{year}{1956}) \bibinfo{pages}{1--23}.
\bibitem[{Winters and Young(2009)}]{Winters2009}
\bibinfo{author}{K.~B. Winters}, \bibinfo{author}{W.~R. Young},
\newblock \bibinfo{title}{Available potential energy and buoyancy variance in
  horizontal convection},
\newblock \bibinfo{journal}{Journal of Fluid Mechanics} \bibinfo{volume}{629}
  (\bibinfo{year}{2009}) \bibinfo{pages}{221--230}.
\bibitem[{Grimmond and Oke(2002)}]{Grimmond2002}
\bibinfo{author}{C.~S.~B. Grimmond}, \bibinfo{author}{T.~R. Oke},
\newblock \bibinfo{title}{Turbulent heat fluxes in urban areas: observations
  and a local-scale urban meteorological parameterization scheme ({LUMPS})},
\newblock \bibinfo{journal}{J. Appl. Meteor} \bibinfo{volume}{41}
  (\bibinfo{year}{2002}) \bibinfo{pages}{792--810}.
\bibitem[{Zhang et~al.(2014)Zhang, Wang, and Peng}]{Zhang2014}
\bibinfo{author}{N.~Zhang}, \bibinfo{author}{X.~Wang},
  \bibinfo{author}{Z.~Peng},
\newblock \bibinfo{title}{Large-eddy simulation of mesoscale circulations
  forced by inhomogeneous urban heat island},
\newblock \bibinfo{journal}{Boundary-Layer Meteorology} \bibinfo{volume}{151}
  (\bibinfo{year}{2014}) \bibinfo{pages}{179--194}.
\bibitem[{Kimura(1975)}]{Kimura1975}
\bibinfo{author}{R.~Kimura},
\newblock \bibinfo{title}{Dynamics of steady convections over heat and cool
  islands},
\newblock \bibinfo{journal}{Journal of the Meteorological Society of Japan}
  \bibinfo{volume}{53} (\bibinfo{year}{1975}) \bibinfo{pages}{440--457}.
\bibitem[{Moeng(1984)}]{Moeng84}
\bibinfo{author}{C.-H. Moeng},
\newblock \bibinfo{title}{A large-eddy-simulation model for the study of
  planetary boundary-layer turbulence},
\newblock \bibinfo{journal}{Journal of the Atmospheric Sciences}
  \bibinfo{volume}{41} (\bibinfo{year}{1984}) \bibinfo{pages}{2052--2062}.
\bibitem[{Mukherjee et~al.(2016)Mukherjee, Schalkwijk, and
  Jonker}]{Mukherjee2016}
\bibinfo{author}{S.~Mukherjee}, \bibinfo{author}{J.~Schalkwijk},
  \bibinfo{author}{H.~J.~J. Jonker},
\newblock \bibinfo{title}{Predictability of dry convective boundary layers: An
  les study},
\newblock \bibinfo{journal}{Journal of the Atmospheric Sciences}
  \bibinfo{volume}{73} (\bibinfo{year}{2016}) \bibinfo{pages}{2715--2727}.
\bibitem[{Smolarkiewicz et~al.(2014)Smolarkiewicz, K\"{u}hnlein, and
  Wedi}]{Piotr2014}
\bibinfo{author}{P.~K. Smolarkiewicz}, \bibinfo{author}{C.~K\"{u}hnlein},
  \bibinfo{author}{N.~P. Wedi},
\newblock \bibinfo{title}{A consistent framework for discrete integrations of
  soundproof and compressible pdes of atmospheric dynamics},
\newblock \bibinfo{journal}{J. Comput. Phys.} \bibinfo{volume}{263}
  (\bibinfo{year}{2014}) \bibinfo{pages}{185--205}.
\bibitem[{Smolarkiewicz et~al.(2017)Smolarkiewicz, K{\"u}hnlein, and
  Grabowski}]{Smolarkiewicz2017}
\bibinfo{author}{P.~K. Smolarkiewicz}, \bibinfo{author}{C.~K{\"u}hnlein},
  \bibinfo{author}{W.~W. Grabowski},
\newblock \bibinfo{title}{A finite-volume module for cloud-resolving
  simulations of global atmospheric flows},
\newblock \bibinfo{journal}{Journal of Computational Physics}
  \bibinfo{volume}{341} (\bibinfo{year}{2017}) \bibinfo{pages}{208 -- 229}.
\bibitem[{Alam and Fitzpatrick(2018)}]{Alam2018}
\bibinfo{author}{J.~M. Alam}, \bibinfo{author}{L.~P.~J. Fitzpatrick},
\newblock \bibinfo{title}{Large eddy simulation of urban boundary layer flows
  using a canopy stress method},
\newblock \bibinfo{journal}{Computers \& Fluids} \bibinfo{volume}{171}
  (\bibinfo{year}{2018}) \bibinfo{pages}{65--78}.
\bibitem[{Wyngaard(2004)}]{Wyngaard2004}
\bibinfo{author}{J.~C. Wyngaard},
\newblock \bibinfo{title}{{Toward Numerical Modeling in the ''Terra
  Incognita''}},
\newblock \bibinfo{journal}{Journal of the atmospheric sciences}
  \bibinfo{volume}{3} (\bibinfo{year}{2004}) \bibinfo{pages}{1816--1826}.
\bibitem[{Kurowski and Teixeira(2018)}]{Kurowski2018}
\bibinfo{author}{M.~J. Kurowski}, \bibinfo{author}{J.~Teixeira},
\newblock \bibinfo{title}{A scale-adaptive turbulent kinetic energy closure for
  the dry convective boundary layer},
\newblock \bibinfo{journal}{Journal of the Atmospheric Sciences}
  \bibinfo{volume}{75} (\bibinfo{year}{2018}) \bibinfo{pages}{675--690}.

\end{thebibliography}

\end{document}